\documentclass[pdflatex, sn-mathphys-num]{sn-jnl}% Math and Physical Sciences Numbered Reference Style 
\usepackage{graphicx}%
\usepackage{multirow}%
\usepackage{amsmath,amssymb,amsfonts}%
\usepackage{mathrsfs}%
\usepackage[title]{appendix}%
\usepackage{textcomp}%
\usepackage{manyfoot}%
\usepackage{booktabs}%
\usepackage{algorithm}%
\usepackage{algorithmicx}%
\usepackage{algpseudocode}%
\usepackage{listings}%
\usepackage{gensymb}%
\usepackage[table,xcdraw]{xcolor}%
\usepackage{siunitx}%
\usepackage{tablefootnote}%
\usepackage{hyperref}
\usepackage{xcolor}
\usepackage{ulem}
\usepackage{geometry}   % For margin control
\usepackage{changepage} % To modify margins temporarily
\usepackage{lineno}

\begin{document}

    \title[Article Title]{Neuromorphic Cameras in Astronomy: Unveiling the Future of Celestial Imaging Beyond Conventional Limits}
    
    % neuromorphic Cameras in Astronomy: Unveiling the Future of Celestial Imaging Beyond Conventional Limits
    % "Beyond Pixels: Harnessing neuromorphic Cameras for Unprecedented Insights into the Cosmos"
    % "neuromorphic Cameras: A New Dawn in Astronomical Observation"
    % "Innovative Eyes on the Sky: neuromorphic Cameras in Astronomy"
    % "Astronomy Redefined: Unleashing neuromorphic Camera Potential"
    
    
\author[1]{\fnm{Satyapreet Singh} \sur{Yadav}}\email{satyapreets@iisc.ac.in}

\author[2]{\fnm{Bikram } \sur{Pradhan}}\email{bikram@istrac.gov.in}

\author[1]{\fnm{Kenil Rajendrabhai} \sur{Ajudiya}}\email{kenilr@iisc.ac.in}

\author[3]{\fnm{T. S.} \sur{Kumar}}\email{kumar@aries.res.in}

\author[1]{\fnm{Nirupam} \sur{Roy}}\email{nroy@iisc.ac.in}

\author[4]{\fnm{Andre Van} \sur{Schaik}}\email{a.vanschaik@westernsydney.edu.au}

\author*[1]{\fnm{Chetan Singh} \sur{Thakur}}\email{csthakur@iisc.ac.in}

\affil*[1]{ \orgname{Indian Institute of Science}, \orgaddress{\country{India}}}

\affil*[2]{ \orgname{Indian Space Research Organization}, \orgaddress{\country{India}}}

\affil*[3]{ \orgname{Aryabhatta Research Institute of observational sciencES}, \orgaddress{\country{India}}}

\affil*[4]{ \orgname{International Centre for Neuromorphic Systems}, \orgaddress{\country{WSU, Australia}}}

\abstract{To deepen our understanding of optical astronomy, we must advance imaging technology to overcome conventional frame-based cameras' limited dynamic range and temporal resolution. Our Perspective paper examines how neuromorphic cameras can effectively address these challenges. Drawing inspiration from the human retina, neuromorphic cameras excel in speed and high dynamic range by utilizing asynchronous pixel operation and logarithmic photocurrent conversion, making them highly effective for celestial imaging. We use 1300 mm terrestrial telescope to demonstrate the neuromorphic camera's ability to simultaneously capture faint and bright celestial sources while preventing saturation effects. We illustrate its photometric capabilities through aperture photometry of a star field with faint stars. Detection of the faint gas cloud structure of the Trapezium cluster during a full moon night highlights the camera's high dynamic range, effectively mitigating static glare from lunar illumination. Our investigations also include detecting meteorite passing near the Moon and Earth, as well as imaging satellites and anthropogenic debris with exceptionally high temporal resolution using a 200mm telescope. Our observations show the immense potential of neuromorphic cameras in advancing astronomical optical imaging and pushing the boundaries of observational astronomy.}


\maketitle
\newgeometry{left=10mm, right=10mm, top=15mm, bottom=15mm}
\section{Introduction}\label{sec1}

The universe has always been a source of knowledge and holds many mysteries that attract human attention. Our understanding of it is constantly evolving with technological advancements. The evolution of observational technology, from elementary tools to more advanced telescopes and cameras, has shaped our ability to understand the cosmos. Innovations like Charge-Coupled Devices (CCD) \cite{howell2000handbook, ratledge2012art, mackay1986charge, lesser2015summary} and scientific complementary metal-oxide semiconductor (sCMOS) \cite{alarcon2023scientific, zienkiewicz2024innovative, qiu2021research,wang2020test} have revolutionized optical astronomy, allowing us to capture intricate planetary details, detect faint stars or galaxies, and improved our ability to observe and analyze celestial objects such as asteroids and supernovae \cite{howell2000handbook, berry_burnell, dick2013pluto}.

Though many breakthroughs in optical astronomy have been achieved through CCD-based technology, CCD is not used during full moon night due to bright sky-background conditions. CCD sensors suffer through saturation effects in the presence of bright sources, thereby constraining their dynamic range, particularly in deep-sky imaging scenarios \cite{catterall1997ccd}. CCD sensors are also susceptible to charge saturation and blooming effects, and while anti-blooming measures are used to mitigate blooming effects, their use can reduce quantum efficiency \cite{krishnamurthy2019precision, neely1993ccd, magnan2003detection}. The limited readout speed of CCD sensors curtails their effectiveness in detecting and characterizing rapidly changing or mobile celestial objects \cite{karpov2021characterization, mu2024astronomical}.

CMOS detectors \cite{nikonov2013overview} have undergone rapid advancements, driven by their cost effectiveness and widespread availability in recent years. They offer distinct advantages over traditional CCD sensors, including high frame rates and generally lower readout noise. Their adoption in astronomy gained traction with the introduction of back-illuminated models \cite{pain2005back, alarcon2023scientific, suntharalingam2007back}. However, both the CCD and the CMOS sensors synchronize the digitization of pixel arrays at a fixed frame rate predetermined before acquisition, which poses limitations when dealing with dynamic processes occurring at varying timescales \cite{cabriel2023event}. Consequently, telescopes often resort to fast photomultipliers to capture dynamic phenomena with intricate details \cite{hoang2023neuromorphic, meddi2012new}. This limitation requires sensors that can capture dynamic astronomical events asynchronously to obtain more accurate and detailed observations.

The handling of the vast amounts of data collected by CCD and CMOS sensors, which lack built-in compression technology, is a challenging task currently faced by many observatories \cite{sen2022astronomical, 7363840, mickaelian2019role, faaique2024overview, sachdeva2023big, zhou2024bioinspired}. For example, the Sloan Digital Sky Survey (SDSS) telescope generates $\sim$ 200 gigabytes of data every night \cite{7887648}. The Large Synoptic Survey Telescope (LSST) is expected to exacerbate this challenge, with an expected nightly data output of $\sim$ 200 petabytes \cite{zhang2015astronomy}. To alleviate this data burden, it is crucial to use imaging sensors that output only relevant data, avoiding storing redundant information such as background intensities and minimizing noise.

The degradation of the night sky due to light pollution and satellite streaks poses a significant threat to ground-based astronomy, particularly with the operation of large-scale telescopic facilities \cite{hainaut2020impact, lawler2021visibility, tyson2020mitigation, kocifaj2021proliferation}. These observatories are at a high risk since capturing faint objects requires longer exposure times and wide-field imaging is performed to detect transient events. A satellite passing through the observatory's field of view (FoV) can saturate the sensor for several minutes, impacting significant observation sessions. The launch of numerous satellites in low-Earth orbits has the potential to severely impact astronomical observations, necessitating the integration of fast-acting wide-field cameras along with high-precision astronomical cameras. The fast-acting camera can quickly detect when a satellite is about to enter the astronomical camera's FoV. Upon detection, the astronomical camera shutter closes and reopens once the satellite has moved out of the FoV, minimizing disruption to ongoing observations.

Neuromorphic cameras offer a promising solution to the challenges facing CCD and CMOS sensors. Inspired by the biological functioning of the eye, neuromorphic cameras represent a paradigm change in visual information acquisition \cite{mahowald1994silicon, gallego2020event, posch2014retinomorphic}. With a high dynamic range of $>$100dB, surpassing standard cameras' 60dB, neuromorphic cameras excel in capturing scenes with diverse brightness levels \cite{4252856, 9156346, messikommer2022multi}. Performing asynchronously, neuromorphic cameras encode only brightness changes per pixel \cite{6906931}, substantially reducing data transmission requirements. These cameras operate without shutter speed or exposure time constraints, sampling light based on scene dynamics rather than fixed intervals \cite{delbruck2008frame}, resulting in an exceptionally high temporal resolution of the order of $\micro$s \cite{lenero20113, 6407468}. These characteristics make neuromorphic cameras particularly well-suited for the fast detection and tracking of rapidly evolving astronomical phenomena.

Recent developments in neuromorphic camera technology have sparked considerable interest in the domain of space science and technology, particularly for applications such as space environment monitoring \cite{cohen2019event, 9142352, mcmahon2021commercial}, atmospheric turbulence characterization \cite{boehrer2019using, polnau2021atmospheric, cohen2022exploring}, adaptive optics \cite{kong2019using, kong2020shack, cohen2022exploring}, orbit determination \cite{wisentaner2022orbit}, space object characterization \cite{jolley2022evaluation, jolley2023neuromorphic, jolley2022characterising}, and spacecraft landings \cite{sikorski2021event}. Efforts have been directed towards real-time tracking of space objects \cite{ralph2022real, felsen_detecting_2022}, both from terrestrial and space-based observations, employing single and binocular cameras \cite{ralph2023shake} with feature-based detection \cite{9142352, afshar2019investigation}, and clustering-based algorithms \cite{dong2023event}. Studies have explored the detection limits of neuromorphic cameras \cite{westerhout2023analysis, mcmahon2021commercial, zolnowski2019observational}, astrometric calibration, and source characterization for space imaging \cite{cohen2018approaches, ralph2023astrometric}. Researchers have investigated using neuromorphic cameras for star tracking due to their higher operating speeds, which help mitigate motion blur effects \cite{chin2019star}. Neuromorphic cameras have also made their way into space, with installations such as Falcon-Neuro on the International Space Station (ISS) for the detection of sprites and lightning \cite{mcharg2022falcon, mcharg2020falcon}. Though these studies are promising and highlight the potential use cases of neuromorphic cameras in space science, a detailed observatory study of the advantages of neuromorphic cameras in the field of optical astronomy has not been done so far.

In our paper, we comprehensively explore the capabilities and applications of neuromorphic camera technology in astronomy. We begin with a detailed explanation of the operational principles behind neuromorphic cameras, followed by an illustration of their advantages using observations of a full moon night sky. We then present the results of our photometric analysis of stars, high-dynamic imaging of star clusters, and tracking of fast-moving objects using terrestrial telescopes in India. We also discuss serendipitous observation of a possible meteorite passing close to the moon's surface. Finally, we discuss the various potential applications of neuromorphic cameras and propose directions for future research in this field.

\section{Neuromorphic Camera: Working and Characteristics}

\begin{figure}[h]
\centering
\includegraphics[width=1\columnwidth]{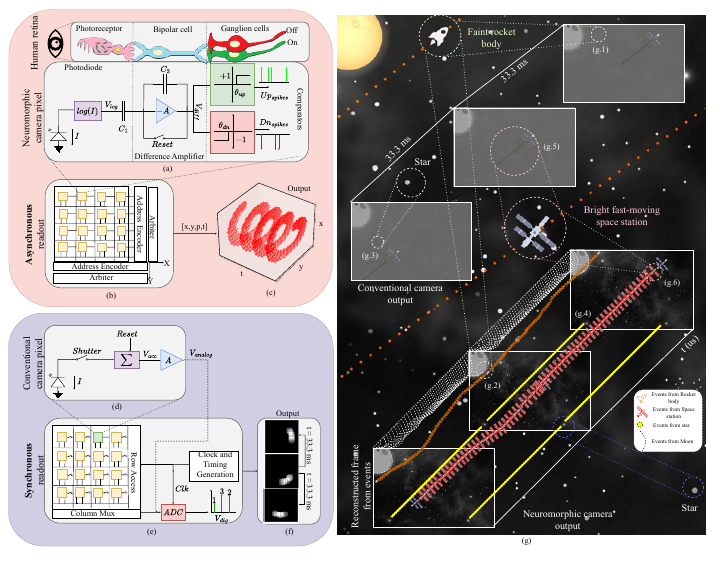}
\caption{Working principles of neuromorphic and conventional cameras. (a) Structure of a neuromorphic camera pixel, with components analogous to the human retina: a photodiode (photoreceptor), difference amplifier (bipolar cells), and comparators (on-off ganglion cells), including the transformation of light intensity into the logarithmic domain and the spike generation process. (b) Asynchronous data transmission in the neuromorphic camera using AER readout. (c) The high temporal resolution of the neuromorphic camera enables tracking of the fast-rotating dot without motion blur. (d) Conventional camera pixel components: photodiode, charge storage, shutter switch, and gain unit. (e) Conventional camera workflow showing digital conversion by the ADC and frame generation. (f) Example of a fast-rotating dot in a static background. Frame generation at 30fps leads to motion blur. (g) Illustration of a full moon night sky showcasing the neuromorphic camera's ability to capture faint objects like rocket debris and dim stars with minimal interference from bright moonlight and accurately tracking the fast-moving space station. In contrast, the conventional camera's output is significantly impacted by intense moonlight and motion blur, obscuring faint objects and blurring the space station.}
\label{working_fig}
\end{figure}

Each pixel in a neuromorphic camera includes a photodiode, a difference amplifier, and comparators, which mimic the functions of photoreceptors, bipolar cells, and on-off ganglion cells in the retina, as presented in Fig.\ref{working_fig}-[a]. The photodiode, analogous to photoreceptor cells, detects light intensity. The difference amplifier, similar to bipolar cells, calculates changes in light intensity. The comparators function like on-off ganglion cells, generating positive and negative events when changes in brightness exceed or fall below a certain contrast threshold \cite{gallego2020event, lichtsteiner200364x64, lichtsteiner200564x64, lichtsteiner2006128}. The photodiode converts incoming light into an electrical current proportional to the light intensity. An amplifier, with transistors operating in the subthreshold region, transforms the photodiode current into a voltage that varies logarithmically with the intensity, as described by the following equation:
\begin{equation*}
V_{\text{log}}(t) = \frac{kT}{q} \ln\left(\frac{I_{\text{ph}}(t)}{I_0}\right)
\end{equation*}
where $V_{\text{log}}$ is the output voltage, $k$ is the Boltzmann constant, $T$ is the temperature (in Kelvin), $q$ is the charge of an electron, $I_0$ is a reference current, and $I_{\text{ph}}(t)$ is the photodiode current.
The output of the difference amplifier can be expressed as:
\begin{equation*}
V_{\text{diff}}(t) = V_{log}(t) - V_{ref}
\end{equation*}
An event $p_{k}$ at time $t = t_{k}$ is generated when
\begin{equation*}
p_{k} = \begin{cases}
+1 & \text{if } V_{\text{diff}}(t = t_{k}) \geq \theta_{\text{up}} \\
-1 & \text{if } V_{\text{diff}}(t = t_{k}) \leq \theta_{\text{dn}}
\end{cases}
\end{equation*}

Where $\theta_{up}, \theta_{dn}$ are the temporal contrast thresholds for increase and decrease in pixel intensity. After each event generation, the reference voltage $V_{ref}$ is updated to the current log voltage, $V_{log}(t)$: $V_{ref}\leftarrow V_{log}(t = t_{k})$.

Each pixel asynchronously relays its $x$ and $y$ coordinates, along with polarity $p_{k}$ and timestamp $t_{k}$, to the camera bus, as depicted in Fig.\ref{working_fig}-[b]. These events traverse from the pixel grid to the output interface using address event representation (AER) readout \cite{boahen2004burst, liu2014event}. The asynchronous nature of this transmission provides high temporal resolution, effectively mitigating motion blur in dynamically changing scenes, such as a fast rotating dot on a screen, as shown in Fig.\ref{working_fig}-[c].

A typical conventional camera pixel consists of a photodiode for light detection, a charge storage component, and additional circuitry such as a shutter switch and gain unit for processing the captured signal (see Fig.\ref{working_fig}-[d]). The shutter switch controls the exposure time, determining how long the sensor is exposed to light. The gain unit adjusts the amplification of the signal captured by the sensor, allowing for brightness adjustment in different lighting conditions. Once the sensor captures the analog signal, the Analog-Digital Converter (ADC) converts it into a digital format for processing and storage (see Fig.\ref{working_fig}-[e]) \cite{skorka2014cmos}. Unlike neuromorphic cameras, which operate asynchronously and transmit event-based data, conventional cameras acquire information at a pre-defined rate specified by an external clock, such as 30 frames per second (fps). Such fixed readout speeds often result in motion blur for fast-moving objects, as depicted in Fig.\ref{working_fig}-[f].

\subsection{Advantages of a Neuromorphic Camera}
An example scenario is depicted in Fig.\ref{working_fig}-[g] to highlight the advantages of the neuromorphic camera. The scene features a night sky with a full moon, accompanied by both bright and faint stars. A faint rocket body passes close to the brightly illuminated moon while a fast-moving space station crosses the camera's FoV.

\subsubsection{High Dynamic Range}
The incident light intensity on a pixel comprises the constant background illumination and the varying brightness of the source. When transformed into the logarithmic domain, these components get added up. As the neuromorphic camera detects contrast changes, it effectively cancels out the constant background illumination, leaving only the variations in intensity due to the source at the output \cite{lichtensteiner2008128x128}. 
This high dynamic range capability allows it to capture faint objects such as rocket debris (Fig. \ref{working_fig}-[g.2]) and dim stars (Fig.\ref{working_fig}-[g.4]) against the bright lunar background. In contrast, conventional frame-based cameras are significantly affected by the intense moonlight, which diminishes the visibility of these faint celestial features (Fig. \ref{working_fig}-[g.1,g.3]).
    
\subsubsection{High Temporal Resolution}
The asynchronous operation of each pixel in a neuromorphic camera, combined with event generation in the analog domain, enables it to capture dynamic scenes effectively. $\micro s$-level precision is achieved through timestamping in the digital domain, facilitated by a MHz clock \cite{gallego2020event, 8946715, Scheerlinck_2020_WACV}. This capability allows the neuromorphic camera to accurately track fast-moving objects, such as a space station in the night sky (Fig.\ref{working_fig}-[g.6]), without experiencing motion blur. In contrast, frame-based conventional cameras, operating at frame rates of 30/60 fps, often encounter motion blur, complicating the localization within the captured frames (Fig.\ref{working_fig}-[g.5]).
    
\subsubsection{Event-driven Data Rate}
The neuromorphic camera operates as an efficient, data-driven sensor by remaining inactive in static scenes where no brightness changes occur and only activating in response to dynamic variations \cite{gallego2020event}. This approach optimizes data storage by transmitting information solely based on scene dynamics. Conventional frame-based cameras capture both static and dynamic information, often leading to large volumes of data generation, resulting in high storage and computational demands.

\subsubsection{Low Power Consumption}
A conventional frame-based camera consumes power due to continuous scene sampling, often leading to higher power consumption, elevating noise, and generating heat, thus requiring cooling mechanisms for longer operating hours. Each pixel in a neuromorphic camera consumes minimal power, typically in the micro-watt range \cite{gallego2020event, 8702508, lenero20113}. Only the parts of the sensor that generate events use power, which makes the overall power use more efficient. This efficiency may reduce the need for extra cooling, a benefit for long astronomical observations.

\subsection{Limitations of a Neuromorphic Camera}
Neuromorphic cameras offer significant advantages in high-speed imaging and dynamic range but present several limitations in astronomical applications. Like CMOS sensors, neuromorphic cameras utilize photodiode technology, with modern designs incorporating 3D stacking to enhance the photodiode area \cite{9063149}. Despite these advancements, their quantum efficiency, peaking at 78\% at 505 nm, remains lower than that of CCD sensors, which typically achieve over 95\% QE \cite{lesser2015summary}, making CCDs the preferred choice for high-sensitivity astronomical observations. Neuromorphic cameras generally have lower pixel densities than traditional CCD and CMOS sensors, limiting their ability to capture fine details and making them less suitable for high-resolution imaging of celestial objects.

Unlike CCD and CMOS sensors, which benefit from well-established calibration techniques, such as bias, dark, and flat-field corrections—neuromorphic sensors lack a standardized calibration approach for astronomical applications. Developing robust correction techniques to address background noise, sensor artifacts, and response non-uniformity remains an ongoing challenge. The number of events generated by a neuromorphic camera is determined by changes in contrast within the scene that exceed the temporal contrast threshold. Consequently, event generation is directly influenced by this threshold, with lower thresholds producing more events compared to higher thresholds. For accurate photometric measurements, it is essential to develop a model that accounts for the impact of the temporal contrast threshold on event generation, ensuring precise data analysis.

\subsection{Noise sources in a Neuromorphic Camera}
The asynchronous operation of a neuromorphic camera introduces specific types of noise, which can be categorized into background activity and hot pixels. Background activity consists of unwanted events arising from thermal noise, dark current, and sensor imperfections. This type of noise becomes more prominent in low-light conditions, where photodiodes struggle to differentiate between genuine scene changes and random pixel fluctuations. Background activity is typically spatially uncorrelated and appears randomly across the sensor. On the other hand, hot pixels are defective pixels that continuously generate spurious events, regardless of actual scene dynamics. These artifacts often result from manufacturing defects, extended sensor use, or elevated temperatures \cite{lv2024denoising, annamalai2024beyond}. Unlike background activity, hot pixels are spatially fixed and can be detected and mitigated through calibration or filtering techniques.

\begin{figure}[h]
    \centering
    \includegraphics[width=1\columnwidth]{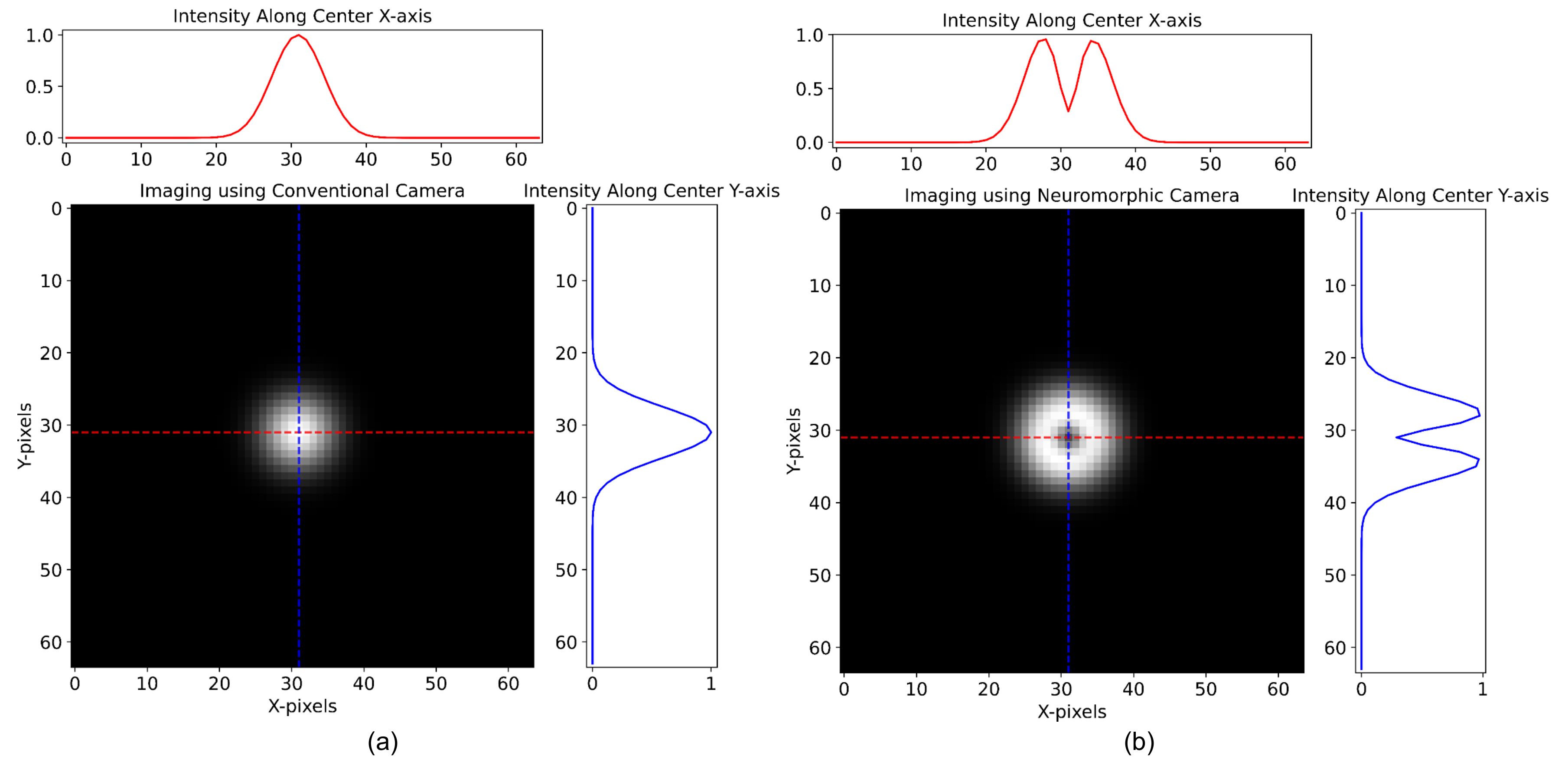}
    \caption{Simulation-based comparison of point source imaging using a conventional and a neuromorphic camera:  
    (a) In a conventional camera, light is integrated over the exposure duration, producing a Gaussian-shaped intensity distribution. The intensity profiles along the horizontal and vertical lines passing through the centre of the star exhibit a Gaussian nature.  
    (b) A neuromorphic camera, in contrast, generates events based on brightness changes caused by atmospheric tip/tilt variations. This results in a doughnut-shaped output after event integration, with distinct intensity distributions along the horizontal and vertical centre lines compared to the conventional camera.}
    \label{image_comparison}
\end{figure}

\subsection{Imaging using a Neuromorphic Camera}
In astronomical imaging, a point source such as a distant star appears as a Gaussian blur in fixed frame rate cameras due to the combined effects of atmospheric turbulence, diffraction, and the optical system's point spread function (PSF), as shown in Fig.\ref{image_comparison}-[a]. In contrast, a neuromorphic camera captures the atmospheric tip-tilt variations of the incoming wavefront as rapid fluctuations in the star's position, generating a continuous stream of distinct temporal events. Rather than averaging these small perturbations over time as traditional sensors do, neuromorphic vision sensors resolve them temporally. Since contrast changes are most pronounced at the edges of the blurred profile, events are primarily registered along the periphery, while the uniform central region produces fewer events. Thus producing the spatial derivative of a conventional Gaussian intensity profile so that the star's reconstructed image assumes a doughnut-like appearance, as shown in Fig.\ref{image_comparison}-[b]. It can be demonstrated that the flux obtained from the doughnut-like profile is directly proportional to that derived from the Gaussian profile.

\section{Observational Setup and Analysis}

The neuromorphic camera was installed on both a large-aperture and a small-aperture telescope to evaluate its performance under different observational conditions. The 1300 mm Devasthal Fast Optical Telescope (DFOT) at the Aryabhatta Research Institute of Observational Sciences (ARIES) in Devasthal, India \cite{joshi2022aries}, equipped with a Cassegrain focal plane, was utilized as the large aperture telescope to evaluate the camera’s high dynamic range by observing bright stars, planets, the Moon, and star clusters. This configuration provided a narrow FoV of $4.16^{\prime} \times 2.34^{\prime} $. A 200 mm Dobsonian telescope in Bangalore was used as a small aperture instrument, offering a wider FoV of $17.8^{\prime} \times 10^{\prime}$. This setup facilitated the analysis of the high temporal resolution characteristics of the camera by observing moving objects, which are examined more effectively with a wider FoV.

\subsection{1300 mm Devasthal Fast Optical Telescope}
\begin{figure}[h]
    \centering
    \includegraphics[width=1\columnwidth]{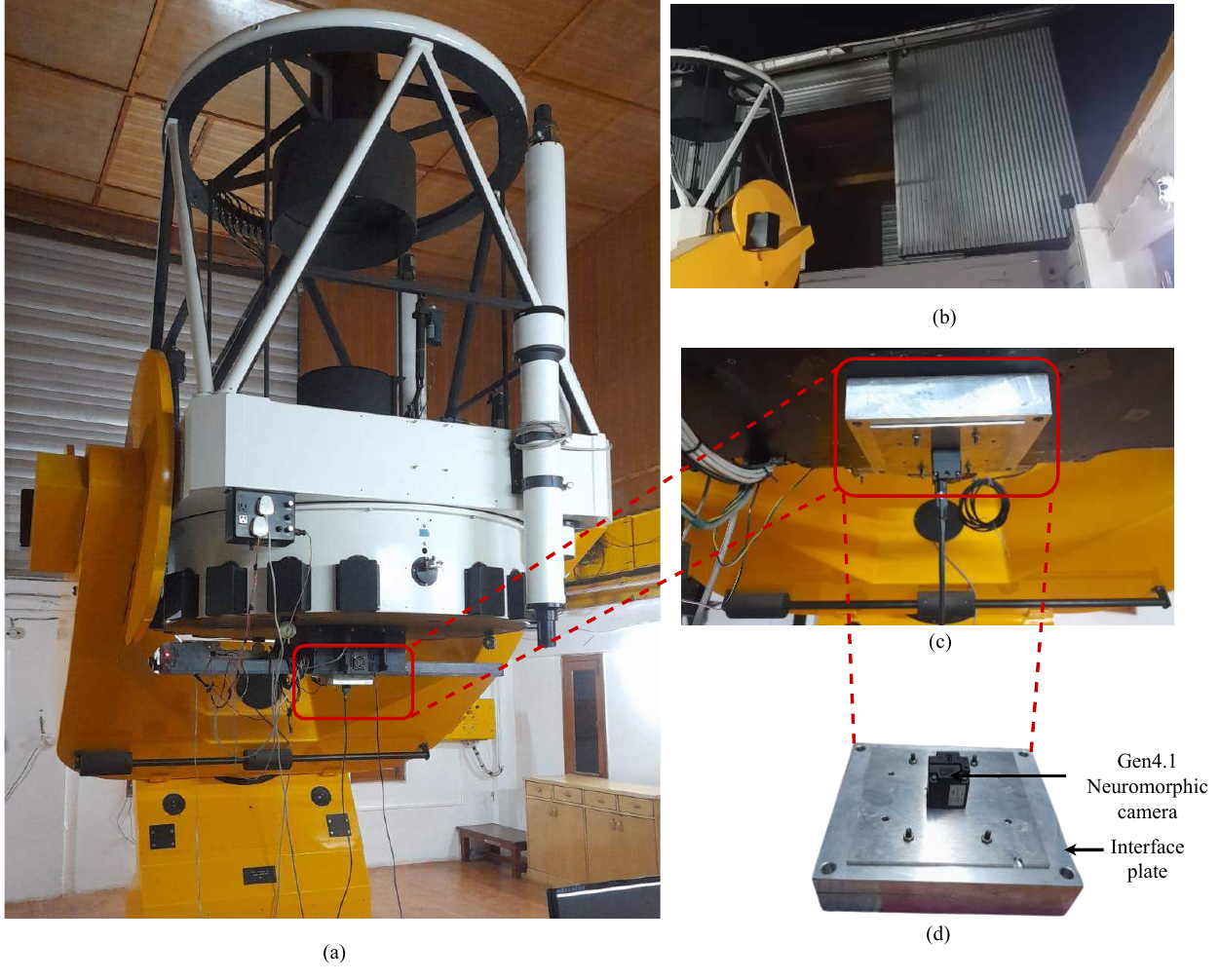}
    \caption{Observational setup for neuromorphic camera-based celestial observations using the 1300 mm DFOT telescope at Devasthal, ARIES, India. (a) The neuromorphic camera installed at the focal plane of the telescope. (b) Roll-off rooftop moving back to expose the telescope to the night sky. (c) Close-up view of the neuromorphic camera mounted on telescope using a custom interface plate. (d) The neuromorphic camera integrated with the interface plate.}
    \label{observational_setup}
\end{figure}

The 1300 mm DFOT adopts a Cassegrain design with a focal ratio of four, offering a broad FoV spanning up to \SI{66}{\arcminute}. With a precision servo-controlled fork equatorial mount, the telescope achieves a pointing accuracy better than \SI{10}{\arcsecond} rms and a tracking accuracy better than 0.5 arc-second/sec for 10 minutes even without its auto guider. Housed within an open roll-off roof structure for effective cooling (see Fig.\ref{observational_setup}-[b]), the telescope has a motorized filter changer capable of accommodating a choice of filters at a time, out of the available broad-band UBVRI, SDSS ugriz, and narrowband H-alpha, O[III], and S[II] interference filters. The Devasthal site, located at a longitude of 79.7$\degree$ E, latitude of 29.4$\degree$ N, and an altitude of $\sim$ 2450 meters above mean sea level, provides exceptional darkness and sub-arcsecond seeing, making it an ideal location for astronomical research.

\subsection{200 mm Dobsonian Telescope}
The 200 mm Dobsonian telescope used in our study is a Newtonian reflector with a 203 mm (8-inch) primary mirror and a focal length of 1200 mm, yielding a focal ratio of f/6. This configuration provides a FoV of $17.8^{\prime} \times 10^{\prime}$. The primary mirror is made from BK7 glass with a surface accuracy of 1/12 wave and features 94\% enhanced reflectivity coatings. The telescope's base incorporates roller bearings for smooth azimuthal movement, facilitating manual tracking of objects across the sky. Observations were conducted in Bangalore, India, at a latitude of approximately 12.97$\degree$ N and a longitude of 77.59$\degree$ E, with an elevation of $\sim$ 900 meters above sea level.

\subsection{Observational Setup}
On the full moon night of November 28, 2023, we carried out observations with the Prophesse Gen4.1 neuromorphic camera \cite{9063149} mounted on the 1300 mm DFOT as seen from Fig.\ref{observational_setup}-[a].  The camera was precisely aligned with the telescope's optical axis using a custom-designed interface plate (see Fig.\ref{observational_setup}-[c, d]). The detailed specifications and quantum efficiency of the Gen4.1 camera are presented in Table \ref{table:camera_specifications}. This setup offered a FoV of $4.16^{\prime} \times 2.34^{\prime} $ with a plate scale of $0.195^{\prime\prime}$ per pixel. The observation plan involved capturing both bright and faint stars on a full moon night, with an observable window ranging from right ascension (RA) $\sim$20h to $\sim$12h and declination (Dec) -30$\degree$ to 90$\degree$, to determine the camera's limiting magnitude. Celestial data were collected using Johnson V, B, and R filters.

For the 200 mm Dobsonian telescope, the neuromorphic camera was interfaced directly using a standard adapter. This setup provided a FoV of $17.8^{\prime} \times 10^{\prime}$ with a plate scale of $0.84^{\prime\prime}$ per pixel. The observation plan focused on capturing fast-moving objects such as debris, satellites, and potential meteoroids.

\begin{table}[h!]
    \centering
    \caption{Gen4.1 Neuromorphic Camera Specifications*}
    \begin{tabular}{|c|c|c|c|c|c|}
        \hline
        \rowcolor[HTML]{DAE8FC} 
        Sensor
        & Resolution
        & \begin{tabular}[c]{@{}c@{}}Pixel pitch \\ (\micro m)\end{tabular}
        & \begin{tabular}[c]{@{}c@{}}Fill factor \\ (\%)\end{tabular}
        & \begin{tabular}[c]{@{}c@{}}Dynamic range \\ (dB)\end{tabular}
        & \begin{tabular}[c]{@{}c@{}}Power \\ consumption (mW)\end{tabular} \\ \hline
        
        Gen4.1      & 1280 x 720        & 4.86      & \textgreater 77       & \textgreater 100      & $\sim$500     \\ \hline
        
        \rowcolor[HTML]{DAE8FC}
        Wavelength      & 455nm     & 505nm     & 625nm     & 850nm     & 940nm     \\ \hline
        
        \begin{tabular}[c]{@{}c@{}}Quantum \\ efficiency (\%)\end{tabular}      & 60        & 78        & 69        & 28        & 13        \\ \hline
    \end{tabular}
    \label{table:camera_specifications}
\footnotesize{*Taken from \cite{9063149} and its datasheet.}
\end{table}

\begin{figure}[h]
    \centering
    \includegraphics[width=1\columnwidth]{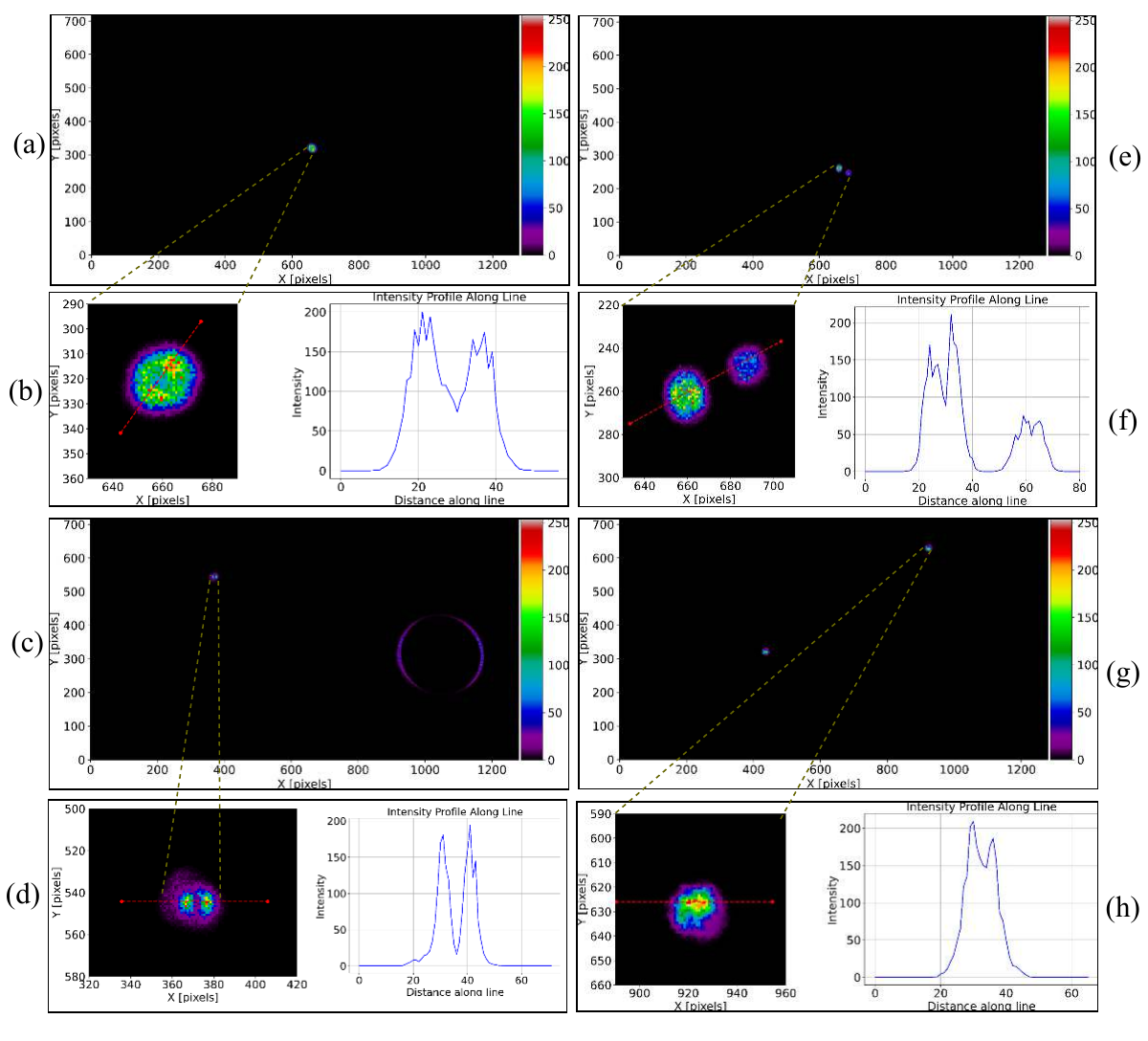}
    \caption{Celestial Observations with the neuromorphic camera on the 1300 mm DFOT. False colors are used to emphasize variations in pixel intensity within the grayscale images for visualization purposes. Images captured include: 
    (a) HIP 9884, V-band magnitude 2, 10s acquisition. (b) Zoomed view of HIP 9884 showing a donut-shaped profile and the corresponding intensity profile along a marked red line. (c) Jupiter and its moon, Ganymede, 10s V-band acquisition. (d) Zoomed view of Ganymede with its intensity profile along a marked red line. (e) The multiple star system SAO 97646, \SI{6}{\arcsecond} separation, 10s V-band acquisition. (f)  Zoomed view of SAO 97646, displaying the intensity profile along the red line. (g) SAO 92721, V-band magnitude 5, 20s acquisition. (h) Zoomed view of a star in SAO 92721 with the corresponding intensity profile along the red line.}
    \label{recordings}
\end{figure}

\subsection{Data Acquisition and Analysis}
After integrating the neuromorphic camera with the 1300 mm DFOT and performing an alignment check, we achieved focus for the test setup. Jupiter was selected as the reference object, and the telescope's focuser was adjusted until the planet appeared as a well-defined disk with minimal variance. This configuration was determined to be the optimal focus setting for observing other celestial bodies. The optimal setting was then programmed into the telescope system, enabling the telescope assembly to maintain stable focus automatically throughout observations. We observed various celestial bodies ranging from stars, planets, moons, and star clusters, as seen in Fig.\ref{recordings} at different camera settings as listed in Table \ref{table:Observation_log}. In the Gen4.1 neuromorphic sensor, bias settings are measured in IDAC units, which indicate the current that controls the sensitivity of the camera's pixels to brightness changes. We fined-tuned the bias settings for each observation based on the target's stellar magnitude and contrast, ensuring that faint objects are detected with high sensitivity while noise is minimized. The IDAC-based design provides stability by compensating for temperature and process variations, which is helpful for astronomical imaging.

\begin{figure}[h]
    \centering
    \includegraphics[width=1\textwidth]{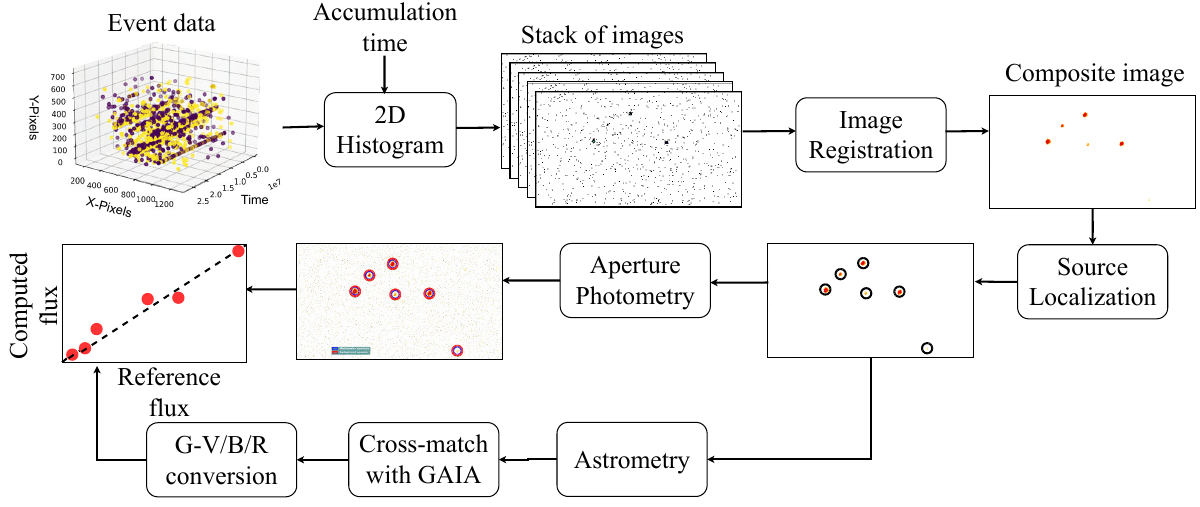}
    \caption{Signal processing steps followed to perform photometric analysis on event data captured from the neuromorphic camera.}
    \label{signal_processing}
\end{figure}

Data were captured in an event-based format represented by tuples $(x, y, p, t)$. We used the 2D histogram technique by aggregating event counts over fixed time windows per pixel. This allowed us to transform the data into grayscale images for astrometry and photometry analysis of the celestial data. This approach enabled us to employ conventional astronomical techniques such as aperture photometry and astrometry, leveraging resources such as the Astropy libraries \cite{astropy:2022} and the GAIA DR3 database \cite{gaia_collaboration_gaia_2021} using the signal processing steps highlighted in Fig.\ref{signal_processing}. 

Although the telescope generally tracked the observed celestial objects, specific recordings, such as those of GD 71 and Trapezium, were captured while the telescope was slewed rather than tracked, resulting in motion drift. To address this, multiple grayscale frames were generated using time binning, and image registration techniques were applied to correct for shifts, producing a high signal-to-noise ratio (SNR) composite image. We then performed source localization to count the stars, followed by aperture photometry and astrometry. To facilitate a comparison between the neuromorphic camera flux and the GAIA database, we converted the G-band flux to the V, B, and R bands using the standard conversion equations outlined in the Gaia Early Data Release 3 documentation, as detailed below, where $G$ is the Gaia G-band magnitude, $G_{BP-RP}$ is the difference between the Gaia blue and red photometric band magnitudes, and $B, V, R$ are the magnitudes in the Johnson-Cousins photometric system.

\begin{align}
G - V &= -0.02704 + 0.01424 \times G_{BP-RP} - 0.2156 \times G_{BP-RP}^2 + 0.01426 \times G_{BP-RP}^3 \\
G - R &= -0.02275 + 0.3961 \times G_{BP-RP} - 0.1243\times G_{BP-RP}^2 - 0.01396\times G_{BP-RP}^3 + 0.003775\times G_{BP-RP}^4 \\
G - V &= -0.04749 - 0.0124 \times (B - V) - 0.2901 \times(B - V)^2 + 0.02008 \times(B - V)^3
\end{align}

\begin{table}[h]
\centering
\caption{Observation Log for 28 November 2023: Objects, Coordinates, and Camera Settings}
\begin{tabular}{|l|l|l|l|l|l|l|}
\hline
\rowcolor[HTML]{DAE8FC} 
\multicolumn{1}{|c|}{\cellcolor[HTML]{DAE8FC}Object}    & \multicolumn{1}{c|}{\cellcolor[HTML]{DAE8FC}\begin{tabular}[c]{@{}c@{}}RA J2000\\  (h m s)\end{tabular}} & \multicolumn{1}{c|}{\cellcolor[HTML]{DAE8FC}\begin{tabular}[c]{@{}c@{}}DEC J2000\\ (° ' ")\end{tabular}} & \multicolumn{1}{c|}{\cellcolor[HTML]{DAE8FC}\begin{tabular}[c]{@{}c@{}}Observation \\ start time \\ (UTC)\end{tabular}} & \multicolumn{1}{c|}{\cellcolor[HTML]{DAE8FC}\begin{tabular}[c]{@{}c@{}}On bias* \\ ($\theta_{up}$)\end{tabular}} & \multicolumn{1}{c|}{\cellcolor[HTML]{DAE8FC}\begin{tabular}[c]{@{}c@{}}Off bias* \\ ($\theta_{dn}$)\end{tabular}} & \begin{tabular}[c]{@{}l@{}}Stellar \\ magnitude \\ (Band)\end{tabular} \\ \hline
HIP 9884                                                & 02 07 10.40                                                                                              & +23 27 44.70                                                                                             & 16:44:31                                                                                                                & 128                                                                                             & 128                                                                                              & 2.01 (V)                                                               \\ \hline
SAO 92721                                               & 01 57 21.058                                                                                             & +17 49 3.20                                                                                              & 17:24:27                                                                                                                & 80                                                                                              & 80                                                                                               & 5.105 (V)                                                              \\ \hline
\begin{tabular}[c]{@{}l@{}}PSR \\ B0136+57\end{tabular} & 01 39 19.744                                                                                             & +58 14 31.73                                                                                             & 18:23:07                                                                                                                & 80                                                                                              & 80                                                                                               & 13.41 (V)                                                              \\ \hline
GD 71                                                   & 05 52 27.620                                                                                             & +15 53 13.23                                                                                             & 18:05:28                                                                                                                & 78                                                                                              & 78                                                                                               & 12.78 (B)                                                              \\ \hline
SAO 97646                                               & 08 12 13.200                                                                                             & +17 38 54.57                                                                                             & 19:52:27                                                                                                                & 100                                                                                             & 100                                                                                              & 5.79 (V)                                                               \\ \hline
Alnitak                                                 & 05 40 45.527                                                                                             & -01 56 33.26                                                                                             & 18:52:23                                                                                                                & 250                                                                                             & 250                                                                                              & 1.77 (V)                                                               \\ \hline
Jupiter                                                 & 02 21 45.4                                                                                               & +12 41 31.7                                                                                              & 16:40:06                                                                                                                & 128                                                                                             & 128                                                                                              & -2.7 (V)                                                                   \\ \hline
Ganymede                                                & 02 21 47.1                                                                                               & +12 42 02.4                                                                                              & 16:40:06                                                                                                                & 128                                                                                             & 128                                                                                              & 4.6 (V)                                                                    \\ \hline
Moon                                                    & 05 27 49.9                                                                                               & +27 26 05.0                                                                                              & 20:08:22                                                                                                                & 100                                                                                             & 100                                                                                              & -12.6 (V)                                                                  \\ \hline
HILTNER 600                                             & 06 45 13.373                                                                                             & +02 08 14.69                                                                                             & 17:55:11                                                                                                                & 78                                                                                              & 78                                                                                               & 10.44 (V)                                                              \\ \hline
TRAPEZIUM                                               & 05 35 16.500                                                                                             & -05 23 13.99                                                                                             & 20:19:32                                                                                                                & 80                                                                                              & 100                                                                                              & 4 (V)                                                                      \\ \hline
\end{tabular}
\label{table:Observation_log}
\footnotesize{*On/Off bias: Temporal contrast thresholds for generating positive and negative events, indicating an increase or decrease in contrast at the pixel.}
\end{table}

\section{Results}
This section presents the results of neuromorphic cameras for astronomical imaging. We begin by examining the photometric response of the neuromorphic camera across V, B, and R bands for a star field that includes HILTNER 600. We then demonstrate its capability to capture bright and faint objects simultaneously by observing Vega, Betelgeuse, and Trapezium star cluster in the Orion Nebula. Additionally, we track fast-moving objects, such as the ISS and space debris, without motion blur using a 200 mm telescope.

\subsection{Photometric Response}
\begin{figure}[h]
    \centering
    \includegraphics[width=1\textwidth]{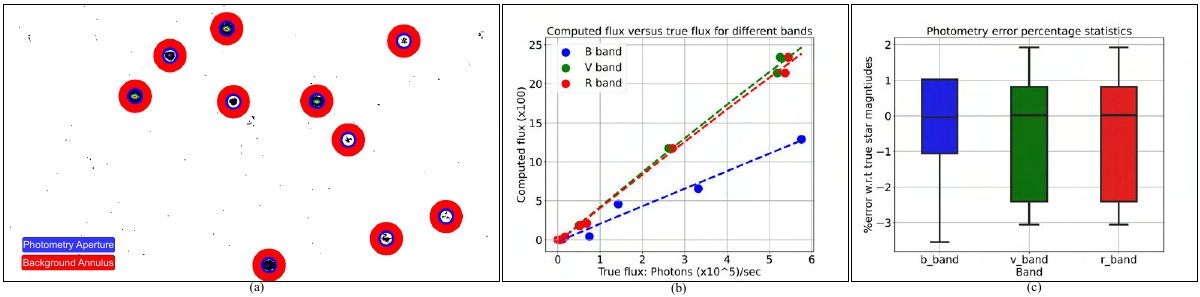}
    \caption{Photometric analysis of the star field containing HILTNER 600: (a) A composite image of the star field containing HILTNER 600 was produced by stacking frames constructed by accumulating events over a 1s duration. Using star localization and aperture photometry techniques, 10 stars were identified within this field. (b) A plot of computed flux using a neuromorphic camera compared against the flux obtained from the GAIA database for Johnson V, B, and R bands, showing a linear relationship. (c) The percentage error in the computed star magnitudes, derived from the flux measurements, was compared to the star magnitudes retrieved from GAIA DR3. }
    \label{photometric_results}
\end{figure}

We captured images of both bright and faint stars using different bias settings. Bright stars were imaged with low sensitivity (high On/Off biases), while faint stars were captured with high sensitivity settings (low On/Off biases). The variation in bias settings complicates the comparison of photometric results across the dataset, as different biases affect the number of events and, consequently, the measured flux. To evaluate the photometric performance of the neuromorphic camera, we conducted photometry on a star field containing the star HILTNER 600. The comparative results, shown in Fig. \ref{photometric_results}, demonstrate a linear relationship for faint stars with V-band magnitudes ranging from 10 to 16.25.  
        
We calculated the star magnitudes using the derived relationship between the computed flux and the true flux from GAIA. We then assessed the error percentage between the computed and true star magnitudes, remaining within 3\%, Fig. \ref{photometric_results}-[c]. Employing a precise calibration technique for the neuromorphic camera will enhance flux estimation accuracy and improve star magnitude calculations.

\subsection{High Dynamic Range Imaging}
\begin{figure}[h]
\centering
\includegraphics[width=1\textwidth]{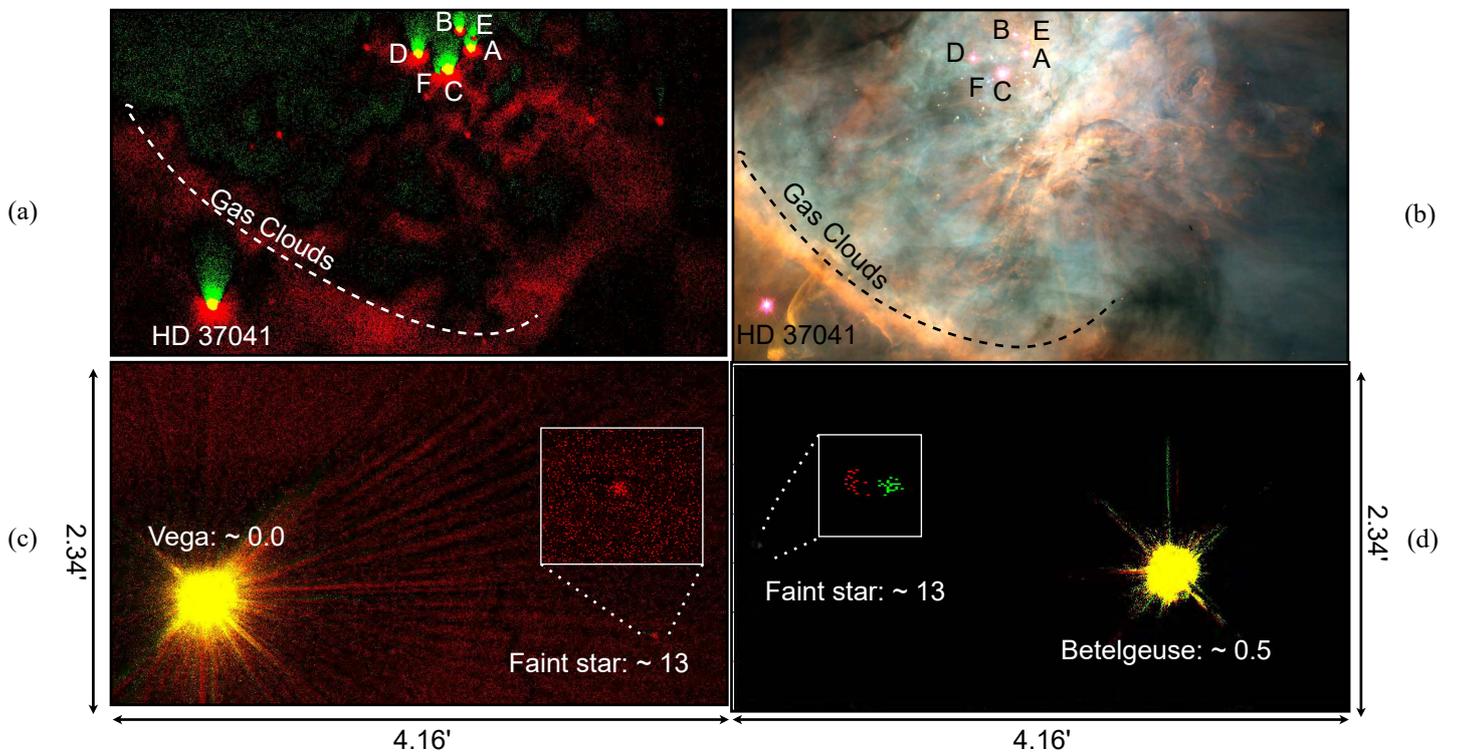}
\caption{High dynamic range of neuromorphic camera:  (a) Image of the Trapezium cluster formed by accumulating events in 200ms window from the neuromorphic camera with a slewing telescope, where red represents positive events and green indicates negative vents. (b) Optical image of the Trapezium star cluster observed through the Hubble Space Telescope \cite{odell1995}. (c) Image of Vega and a nearby faint star, $\sim 200^{\prime\prime}$ apart, demonstrating a dynamic range exceeding 100dB. (d) High dynamic imaging of star Betelgeuse, with a neighbouring faint star visible at roughly $\sim 170^{\prime\prime}$ distance.}
\label{hdr_results}
\end{figure}

The Trapezium star cluster is a challenging target due to its blend of bright stars, faint stars, and even fainter dust cloud structures. With a CCD camera on a moderate-sized telescope, imaging these faint features usually requires several minutes of integration time, often leading to saturation of the brighter stars. Our observations addressed this challenge by slewing the telescope using a neuromorphic camera. This approach takes full advantage of the camera's high dynamic range and temporal resolution, enabling us to capture dust clouds and stars with V-band magnitudes between 5.3 and 14.2 (see Fig. \ref{hdr_results}-[a]), even against the bright background of a full moon night. The gas cloud structures in our image show a clear correlation with those seen in Hubble Space Telescope optical images, Fig. \ref{hdr_results}-[b]. We generated the neuromorphic image by accumulating events over a 200 ms window and applied false color mapping, with red representing positive events and green representing negative events. 
       
Slewing the telescope enhances contrast by creating subtle variations in brightness, which, when combined with atmospheric tip-tilt effects, helps to reveal faint gas clouds in the Trapezium region more clearly. This dynamic contrast change allows for the faster detection of faint objects, a capability that fixed frame rate cameras lack. Fixed frame rate cameras suffer from motion blur during slewing, whereas neuromorphic cameras can capture rapid variations, increasing the likelihood of detecting subtle features. A neuromorphic camera offers a dynamic range exceeding 100 dB, which can simultaneously capture bright and faint sources in a scene. Very bright stars such as Vega, of $\sim$ zero magnitudes and Betelgeuse, of $\sim$ 0.5 magnitudes, are surrounded by much fainter stars. With their limited dynamic range, conventional cameras struggle to detect these faint stars when they lie within about $4^{\prime}$ of the bright stars due to a brighter background. Our observations in tracking mode allowed us to detect faint stars as close as approximately $200^{\prime\prime}$ from Vega and $170^{\prime\prime}$ from Betelgeuse. Since astrometry becomes challenging when only two stars are present in the FoV, we searched within these defined radii for the brightest stars. However, the brightest stars we identified in these regions were magnitude 13, yielding a flux ratio of the order of $10^{5}$, confirming a dynamic range greater than 100 dB.
            
\subsection{High Temporal Resolution Results} 
Moving objects such as satellites, debris, and meteoroids exhibit variable brightness levels in their orbits, which poses challenges for conventional frame-based cameras with fixed temporal resolution. Faint and fast-moving objects often result in motion blur, and reducing the integration time to mitigate this effect requires increasing the camera gain, leading to higher sensor noise levels. In contrast, neuromorphic cameras excel at capturing moving objects by detecting positive and negative events as they traverse the sensor. Our study integrated a neuromorphic camera onto a 200 mm Dobsonian telescope and deployed it over Bengaluru (latitude 12.9716$\degree$ N, longitude 77.5946$\degree$ E, altitude $\sim$ 920 meters above mean sea level) to capture high temporal resolution of satellite imaging and potential meteorite detections.

\subsubsection{Satellite Imaging}
\begin{figure}[h]
\centering
\includegraphics[width=1\textwidth]{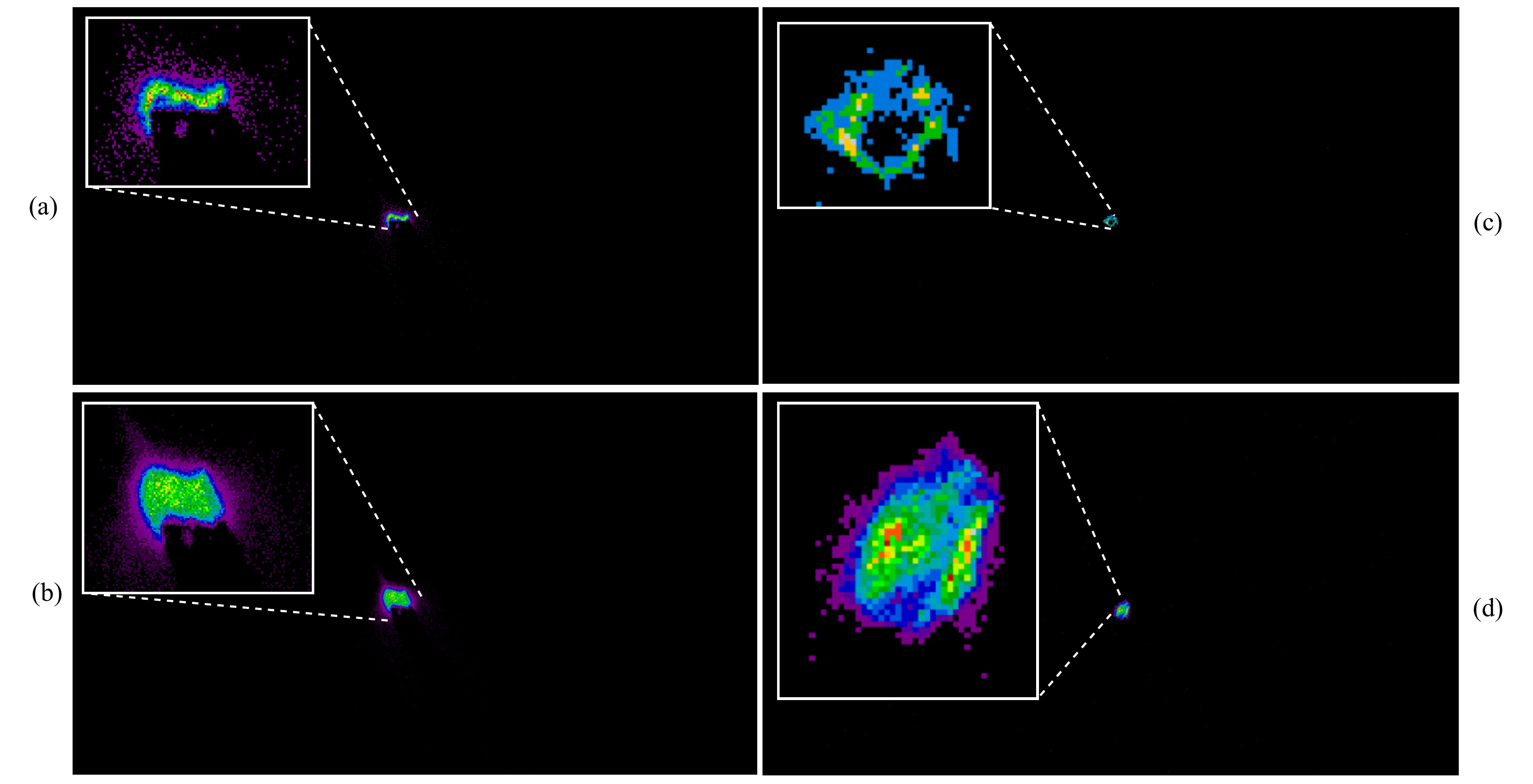}
\caption{The high temporal resolution of the neuromorphic camera is demonstrated through two observational cases. (a) and (b) show the ISS imaged at 23:38:42 UTC on April 14, 2024, with brightness levels ranging from -2.5 to 2.2. Reconstructions of the ISS were generated using positive events, with images obtained at 500 fps in (a) and at 60 fps in (b); note the evident motion blur in the 60 fps image (b). Similarly, the Atlas Centaur-2 R/B was observed at 00:01:46 UTC on April 15, 2024 with brightness levels between 3.4 and 5.2. For this object, reconstructions based solely on positive events were obtained at 500 fps in (c) and at 60 fps in (d), with pronounced motion blur visible in the 60 fps reconstruction (d).}
\label{satellite_results}
\end{figure}

Using a neuromorphic camera, we successfully imaged the International Space Station (ISS) and the Atlas Centaur-2 Rocket Body (R/B) without motion blur by tracking their motion across our FoV (see Fig. \ref{satellite_results}). Due to suboptimal seeing conditions in Bengaluru, we could not capture extremely faint celestial bodies. Moreover, the neuromorphic camera's pixel design causes negative events to be slower than positive events, resulting in their prolonged presence in the frame. To accurately track these fast-moving objects, we reconstructed the event data using only positive events at two integration times: 2ms (equivalent to 500fps), which effectively minimized motion blur, and 16.67ms (equivalent to CCD's 60 fps), which resulted in noticeable motion blur effects.

\subsection{Serendipitous Observations of Moon Meteorites}  
Meteorite impacts on the Moon's surface are common phenomena, appearing as flashes of light when observed through a telescope from Earth. The rates of meteorite impact on the lunar surface are uncertain. They are studied extensively to understand the distribution of meteoroid sizes, which is crucial for evaluating the threat to Earth and spacecraft \cite{nasa_lunar_impact_2024, cudnik2003observation}. The conventional method of studying lunar impacts involves pointing a telescope at the dark side of the Moon and continuously recording video at 30 fps. This video is later processed to detect flashes on the Moon's surface, which typically last less than 0.5 seconds and have a typical travelling speed of 20-70 km/s \cite{pokorny2019meteoroids}.

\begin{figure}[h]
\centering
\includegraphics[width=1\textwidth]{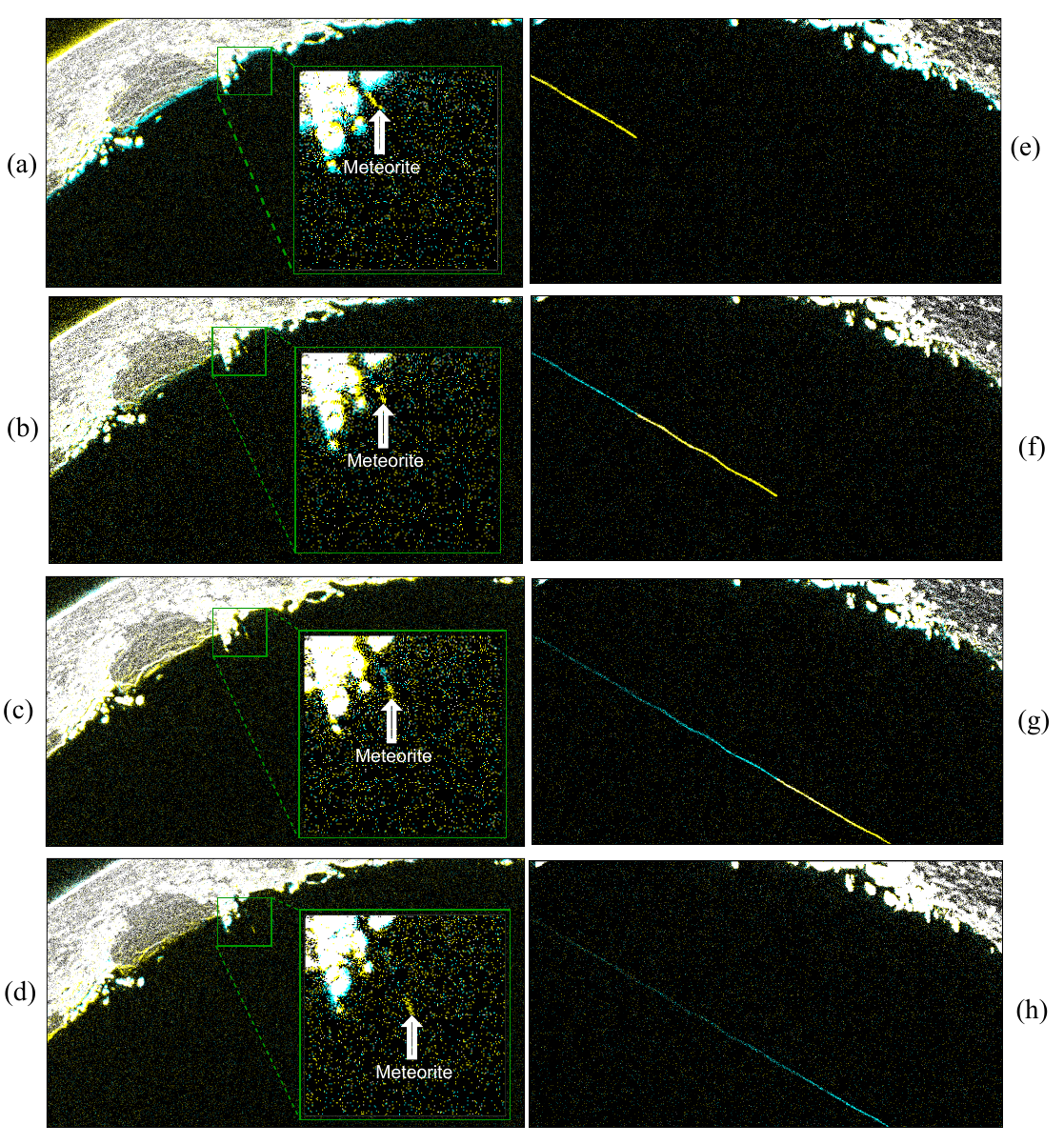}
\caption{Meteorite Snapshots: Snapshots of a potential meteorite passing over the Moon were taken on February 12, 2024, at 13:52:01 UTC. Panels (a), (b), (c), and (d) depict the sequence of the meteorite's trajectory near the Moon. Additional snapshots were taken on February 13, 2024, at 14:21:01 UTC, capturing a potential meteorite passing close to Earth. Panels (e), (f), (g), and (h) illustrate this sequence. Each frame was formed by accumulating events over a 400 ms interval, with positive events displayed in yellow and negative events in blue.}
\label{moon_results}
\end{figure}

We conducted lunar observations using the Gen4.1 neuromorphic camera mounted on a 200 mm Dobsonian telescope in Bengaluru. During these sessions, we serendipitously detected light trails traversing the camera's FoV. To determine whether these trails were due to meteoroids impacting the Moon, Earth-grazing meteors, or satellites orbiting the Moon, we analyzed the angular speeds of the objects. Objects close to Earth would exhibit higher angular velocities compared to those near the Moon. As shown in Fig. \ref{moon_results} (a-d), the average angular velocity of these objects was $\sim$ 50 arc-second/s, corresponding to a linear velocity of about 100 km/s near the lunar surface, suggesting that they could be meteoroids passing near the Moon. In contrast, the objects depicted in Fig. \ref{moon_results} (e-h) exhibited significantly higher angular velocities, indicating they might be meteoroids passing closer to Earth's surface.

\section{Discussions}  
The neuromorphic camera functions asynchronously at each pixel and processes data on a logarithmic scale, delivering high temporal resolution and dynamic range. Our observations employed a moderately large (1300 mm) and a small aperture (200 mm) telescope to capture a broad spectrum of celestial objects, including bright stars, faint entities, and objects with significant apparent motion. We successfully imaged the faint gas clouds of the Trapezium star cluster with a 200 ms accumulation time on a full moon night. We tracked fast-moving satellites and debris using a 2 ms accumulation time without motion blur.
The neuromorphic camera provides an event-driven data rate that is adjustable through On/Off bias settings, offering flexibility in data acquisition. Its compact form factor and low power consumption make it suitable for long astronomical observations, without extra cooling hardware. This contrasts with CCD cameras, which require 60-80W of power for cooling to manage noise levels.

We demonstrated the feasibility of performing photometry and astrometry using established astronomical techniques by converting event data into frames. Our photometric measurements showed a linear relationship for faint objects with V-band magnitudes ranging from 10 to 16.25.
While we employed conventional photometric tools, standard calibrations such as bias, dark, and flat frames which are typically used with frame-based cameras were not applied. The principle of operation of neuromorphic camera is different from conventional sensors and necessitates different calibration approach for accurate data analysis.

The neuromorphic camera offers significant potential for a wide range of astronomical applications. Our results, including the unexpected observation of a meteoroid passing close to the lunar surface, highlight that neuromorphic camera excels in capturing transient and dynamic events.
This capability is particularly valuable for studying binary stars, supernovae, asteroids and comets, precisely observing Jupiter's moon occultations, and investigating meteorite impacts. Neuromorphic cameras can be used to support adaptive optics for atmospheric correction and enhance speckle imaging techniques. Also, it is well-suited for accurate timing measurements of pulsars and variable stars.

CCD cameras provide high-resolution images of static objects, while neuromorphic cameras excel in capturing transient events. Integrating data from both sensors can provide high contrast imaging capabilities. The neuromorphic camera can also effectively address the challenge of satellite interference in night sky observations due to its high temporal resolution. Its ability to detect and track fast-moving objects enables it to identify approaching satellites within its FoV.  Provided that a neuromorphic camera is developed with a sufficiently wide FoV, it can quickly detect an approaching satellite, signaling closing of shutter for conventional camera, thus preventing data corruption and enhancing the reliability of continuous night sky imaging.

The neuromorphic camera is a cutting-edge technology with significant, yet largely unexplored, potential in astronomy. While it has proven its worth in fields such as robotics and automated vehicles, our preliminary findings highlight its promising applications for astronomical research. The camera's ability to capture dynamic and transient celestial phenomena suggests it could offer unique advantages. We hope to inspire the astronomy community to further investigate and adopt this technology by presenting these initial results. Future efforts should focus on developing effective calibration techniques and specialized algorithms to improve data analysis. As research progresses, neuromorphic cameras could transform our ability to observe and interpret dynamic astronomical events, leading to deeper and more detailed insights into the cosmos.

\bibliography{sn-article} 


\begin{thebibliography}{89}
\ifx \bisbn   \undefined \def \bisbn  #1{ISBN #1}\fi
\ifx \binits  \undefined \def \binits#1{#1}\fi
\ifx \bauthor  \undefined \def \bauthor#1{#1}\fi
\ifx \batitle  \undefined \def \batitle#1{#1}\fi
\ifx \bjtitle  \undefined \def \bjtitle#1{#1}\fi
\ifx \bvolume  \undefined \def \bvolume#1{\textbf{#1}}\fi
\ifx \byear  \undefined \def \byear#1{#1}\fi
\ifx \bissue  \undefined \def \bissue#1{#1}\fi
\ifx \bfpage  \undefined \def \bfpage#1{#1}\fi
\ifx \blpage  \undefined \def \blpage #1{#1}\fi
\ifx \burl  \undefined \def \burl#1{\textsf{#1}}\fi
\ifx \doiurl  \undefined \def \doiurl#1{\url{https://doi.org/#1}}\fi
\ifx \betal  \undefined \def \betal{\textit{et al.}}\fi
\ifx \binstitute  \undefined \def \binstitute#1{#1}\fi
\ifx \binstitutionaled  \undefined \def \binstitutionaled#1{#1}\fi
\ifx \bctitle  \undefined \def \bctitle#1{#1}\fi
\ifx \beditor  \undefined \def \beditor#1{#1}\fi
\ifx \bpublisher  \undefined \def \bpublisher#1{#1}\fi
\ifx \bbtitle  \undefined \def \bbtitle#1{#1}\fi
\ifx \bedition  \undefined \def \bedition#1{#1}\fi
\ifx \bseriesno  \undefined \def \bseriesno#1{#1}\fi
\ifx \blocation  \undefined \def \blocation#1{#1}\fi
\ifx \bsertitle  \undefined \def \bsertitle#1{#1}\fi
\ifx \bsnm \undefined \def \bsnm#1{#1}\fi
\ifx \bsuffix \undefined \def \bsuffix#1{#1}\fi
\ifx \bparticle \undefined \def \bparticle#1{#1}\fi
\ifx \barticle \undefined \def \barticle#1{#1}\fi
\bibcommenthead
\ifx \bconfdate \undefined \def \bconfdate #1{#1}\fi
\ifx \botherref \undefined \def \botherref #1{#1}\fi
\ifx \url \undefined \def \url#1{\textsf{#1}}\fi
\ifx \bchapter \undefined \def \bchapter#1{#1}\fi
\ifx \bbook \undefined \def \bbook#1{#1}\fi
\ifx \bcomment \undefined \def \bcomment#1{#1}\fi
\ifx \oauthor \undefined \def \oauthor#1{#1}\fi
\ifx \citeauthoryear \undefined \def \citeauthoryear#1{#1}\fi
\ifx \endbibitem  \undefined \def \endbibitem {}\fi
\ifx \bconflocation  \undefined \def \bconflocation#1{#1}\fi
\ifx \arxivurl  \undefined \def \arxivurl#1{\textsf{#1}}\fi
\csname PreBibitemsHook\endcsname

%%% 1
\bibitem[\protect\citeauthoryear{Howell}{2000}]{howell2000handbook}
\begin{bbook}
\bauthor{\bsnm{Howell}, \binits{S.B.}}:
\bbtitle{Handbook of CCD Astronomy}
vol. \bseriesno{2},
\bedition{2nd} edn.
\bpublisher{Cambridge University Press},
\blocation{Cambridge}
(\byear{2000})
\end{bbook}
\endbibitem

%%% 2
\bibitem[\protect\citeauthoryear{Ratledge}{2012}]{ratledge2012art}
\begin{bbook}
\bauthor{\bsnm{Ratledge}, \binits{D.}}:
\bbtitle{The Art and Science of CCD Astronomy}.
\bpublisher{Springer},
\blocation{New York, NY}
(\byear{2012})
\end{bbook}
\endbibitem

%%% 3
\bibitem[\protect\citeauthoryear{Mackay}{1986}]{mackay1986charge}
\begin{barticle}
\bauthor{\bsnm{Mackay}, \binits{C.D.}}:
\batitle{Charge-coupled devices in astronomy}.
\bjtitle{Annual review of astronomy and astrophysics}
\bvolume{24}(\bissue{1}),
\bfpage{255}--\blpage{283}
(\byear{1986})
\end{barticle}
\endbibitem

%%% 4
\bibitem[\protect\citeauthoryear{Lesser}{2015}]{lesser2015summary}
\begin{barticle}
\bauthor{\bsnm{Lesser}, \binits{M.}}:
\batitle{A summary of charge-coupled devices for astronomy}.
\bjtitle{Publications of the Astronomical Society of the Pacific}
\bvolume{127}(\bissue{957}),
\bfpage{1097}--\blpage{1104}
(\byear{2015})
\end{barticle}
\endbibitem

%%% 5
\bibitem[\protect\citeauthoryear{Alarcon et~al.}{2023}]{alarcon2023scientific}
\begin{barticle}
\bauthor{\bsnm{Alarcon}, \binits{M.R.}},
\bauthor{\bsnm{Licandro}, \binits{J.}},
\bauthor{\bsnm{Serra-Ricart}, \binits{M.}},
\bauthor{\bsnm{Joven}, \binits{E.}},
\bauthor{\bsnm{Gaitan}, \binits{V.}},
\bauthor{\bsnm{Sousa}, \binits{R.}}:
\batitle{Scientific cmos sensors in astronomy: Imx455 and imx411}.
\bjtitle{Publications of the Astronomical Society of the Pacific}
\bvolume{135}(\bissue{1047}),
\bfpage{055001}
(\byear{2023})
\end{barticle}
\endbibitem

%%% 6
\bibitem[\protect\citeauthoryear{Zienkiewicz
  et~al.}{2024}]{zienkiewicz2024innovative}
\begin{botherref}
\oauthor{\bsnm{Zienkiewicz}, \binits{P.}},
\oauthor{\bsnm{Karpi{\'n}ska}, \binits{K.}},
\oauthor{\bsnm{Jamro{\.z}y}, \binits{M.}},
\oauthor{\bsnm{Juszczyk}, \binits{B.}},
\oauthor{\bsnm{Pochapskyi}, \binits{D.}},
\oauthor{\bsnm{Przedpe{\l}ski}, \binits{T.}},
\oauthor{\bsnm{{\L}ukasiewicz}, \binits{J.}},
\oauthor{\bsnm{Czortek}, \binits{N.}},
\oauthor{\bsnm{Brona}, \binits{G.}}:
An innovative scmos based autonomous astronomical camera dedicated to universal
  use for sst and other fields of optical astronomy.
International Journal of Electronics and Telecommunications,
261--266
(2024)
\end{botherref}
\endbibitem

%%% 7
\bibitem[\protect\citeauthoryear{Qiu et~al.}{2021}]{qiu2021research}
\begin{barticle}
\bauthor{\bsnm{Qiu}, \binits{P.}},
\bauthor{\bsnm{Zhao}, \binits{Y.}},
\bauthor{\bsnm{Zheng}, \binits{J.}},
\bauthor{\bsnm{Wang}, \binits{J.-F.}},
\bauthor{\bsnm{Jiang}, \binits{X.-J.}}:
\batitle{Research on performances of back-illuminated scientific cmos for
  astronomical observations}.
\bjtitle{Research in Astronomy and Astrophysics}
\bvolume{21}(\bissue{10}),
\bfpage{268}
(\byear{2021})
\end{barticle}
\endbibitem

%%% 8
\bibitem[\protect\citeauthoryear{Wang et~al.}{2020}]{wang2020test}
\begin{bchapter}
\bauthor{\bsnm{Wang}, \binits{S.}},
\bauthor{\bsnm{Ping}, \binits{Y.}},
\bauthor{\bsnm{Men}, \binits{J.}},
\bauthor{\bsnm{Zhang}, \binits{C.}},
\bauthor{\bsnm{Zhao}, \binits{C.}}:
\bctitle{The test of the 4k scmos camera for astronomical application}.
In: \bbtitle{SPIE Future Sensing Technologies},
vol. \bseriesno{11525},
pp. \bfpage{557}--\blpage{570}
(\byear{2020}).
\bcomment{SPIE}
\end{bchapter}
\endbibitem

%%% 9
\bibitem[\protect\citeauthoryear{Berry and Burnell}{2005}]{berry_burnell}
\begin{bbook}
\bauthor{\bsnm{Berry}, \binits{R.}},
\bauthor{\bsnm{Burnell}, \binits{J.}}:
\bbtitle{The Handbook of Astronomical Image Processing},
\bedition{2nd} edn.
\bpublisher{Willmann-Bell},
\blocation{Richmond, Virginia}
(\byear{2005})
\end{bbook}
\endbibitem

%%% 10
\bibitem[\protect\citeauthoryear{Dick}{2013}]{dick2013pluto}
\begin{bchapter}
\bauthor{\bsnm{Dick}, \binits{S.J.}}:
\bctitle{The pluto affair}.
In: \beditor{\bsnm{Dick}, \binits{S.J.}} (ed.)
\bbtitle{Discovery and Classification in Astronomy: Controversy and Consensus},
pp. \bfpage{9}--\blpage{30}.
\bpublisher{Cambridge University Press},
\blocation{Cambridge}
(\byear{2013})
\end{bchapter}
\endbibitem

%%% 11
\bibitem[\protect\citeauthoryear{Catterall}{1997}]{catterall1997ccd}
\begin{bchapter}
\bauthor{\bsnm{Catterall}, \binits{A.}}:
\bctitle{Ccd imaging from the city}.
In: \bbtitle{The Art and Science of CCD Astronomy},
pp. \bfpage{101}--\blpage{110}.
\bpublisher{Springer},
\blocation{New York}
(\byear{1997})
\end{bchapter}
\endbibitem

%%% 12
\bibitem[\protect\citeauthoryear{Krishnamurthy
  et~al.}{2019}]{krishnamurthy2019precision}
\begin{barticle}
\bauthor{\bsnm{Krishnamurthy}, \binits{A.}},
\bauthor{\bsnm{Villasenor}, \binits{J.}},
\bauthor{\bsnm{Seager}, \binits{S.}},
\bauthor{\bsnm{Ricker}, \binits{G.}},
\bauthor{\bsnm{Vanderspek}, \binits{R.}}:
\batitle{Precision characterization of the tess ccd detectors: Quantum
  efficiency, charge blooming and undershoot effects}.
\bjtitle{Acta Astronautica}
\bvolume{160},
\bfpage{46}--\blpage{55}
(\byear{2019})
\end{barticle}
\endbibitem

%%% 13
\bibitem[\protect\citeauthoryear{Neely and Janesick}{1993}]{neely1993ccd}
\begin{barticle}
\bauthor{\bsnm{Neely}, \binits{A.W.}},
\bauthor{\bsnm{Janesick}, \binits{J.R.}}:
\batitle{A ccd anti-blooming technique for use in photometry}.
\bjtitle{Publications of the Astronomical Society of the Pacific}
\bvolume{105}(\bissue{693}),
\bfpage{1330}
(\byear{1993})
\end{barticle}
\endbibitem

%%% 14
\bibitem[\protect\citeauthoryear{Magnan}{2003}]{magnan2003detection}
\begin{barticle}
\bauthor{\bsnm{Magnan}, \binits{P.}}:
\batitle{Detection of visible photons in ccd and cmos: A comparative view}.
\bjtitle{Nuclear Instruments and Methods in Physics Research Section A:
  Accelerators, Spectrometers, Detectors and Associated Equipment}
\bvolume{504}(\bissue{1-3}),
\bfpage{199}--\blpage{212}
(\byear{2003})
\end{barticle}
\endbibitem

%%% 15
\bibitem[\protect\citeauthoryear{Karpov
  et~al.}{2021}]{karpov2021characterization}
\begin{bchapter}
\bauthor{\bsnm{Karpov}, \binits{S.}},
\bauthor{\bsnm{Christov}, \binits{A.}},
\bauthor{\bsnm{Bajat}, \binits{A.}},
\bauthor{\bsnm{Cunniffe}, \binits{R.}},
\bauthor{\bsnm{Prouza}, \binits{M.}}:
\bctitle{Characterization of modern ccd and cmos sensors for sky surveys}.
In: \bbtitle{Revista Mexicana de Astronomia Y Astrofisica Conference Series},
vol. \bseriesno{53},
pp. \bfpage{190}--\blpage{197}
(\byear{2021})
\end{bchapter}
\endbibitem

%%% 16
\bibitem[\protect\citeauthoryear{Mu et~al.}{2024}]{mu2024astronomical}
\begin{botherref}
\oauthor{\bsnm{Mu}, \binits{H.}},
\oauthor{\bsnm{Fan}, \binits{Z.}},
\oauthor{\bsnm{Zhu}, \binits{Y.}},
\oauthor{\bsnm{Zhang}, \binits{Y.}},
\oauthor{\bsnm{Wu}, \binits{H.}}:
Astronomical test with cmos on the 60-cm telescope at the xinglong observatory,
  naoc.
Research in Astronomy and Astrophysics
(2024)
\end{botherref}
\endbibitem

%%% 17
\bibitem[\protect\citeauthoryear{Nikonov and Young}{2013}]{nikonov2013overview}
\begin{barticle}
\bauthor{\bsnm{Nikonov}, \binits{D.E.}},
\bauthor{\bsnm{Young}, \binits{I.A.}}:
\batitle{Overview of beyond-cmos devices and a uniform methodology for their
  benchmarking}.
\bjtitle{Proceedings of the IEEE}
\bvolume{101}(\bissue{12}),
\bfpage{2498}--\blpage{2533}
(\byear{2013})
\end{barticle}
\endbibitem

%%% 18
\bibitem[\protect\citeauthoryear{Pain et~al.}{2005}]{pain2005back}
\begin{botherref}
\oauthor{\bsnm{Pain}, \binits{B.}},
\oauthor{\bsnm{Cunningham}, \binits{T.}},
\oauthor{\bsnm{Nikzad}, \binits{S.}},
\oauthor{\bsnm{Hoenk}, \binits{M.}},
\oauthor{\bsnm{Jones}, \binits{T.}},
\oauthor{\bsnm{Wrigley}, \binits{C.}},
\oauthor{\bsnm{Hancock}, \binits{B.}}:
A back-illuminated megapixel cmos image sensor
(2005)
\end{botherref}
\endbibitem

%%% 19
\bibitem[\protect\citeauthoryear{Suntharalingam
  et~al.}{2007}]{suntharalingam2007back}
\begin{bchapter}
\bauthor{\bsnm{Suntharalingam}, \binits{V.}},
\bauthor{\bsnm{Rathman}, \binits{D.D.}},
\bauthor{\bsnm{Prigozhin}, \binits{G.}},
\bauthor{\bsnm{Kissel}, \binits{S.}},
\bauthor{\bsnm{Bautz}, \binits{M.}}:
\bctitle{Back-illuminated three-dimensionally integrated cmos image sensors for
  scientific applications}.
In: \bbtitle{Focal Plane Arrays for Space Telescopes III},
vol. \bseriesno{6690},
pp. \bfpage{80}--\blpage{89}
(\byear{2007}).
\bcomment{SPIE}
\end{bchapter}
\endbibitem

%%% 20
\bibitem[\protect\citeauthoryear{Cabriel et~al.}{2023}]{cabriel2023event}
\begin{barticle}
\bauthor{\bsnm{Cabriel}, \binits{C.}},
\bauthor{\bsnm{Monfort}, \binits{T.}},
\bauthor{\bsnm{Specht}, \binits{C.G.}},
\bauthor{\bsnm{Izeddin}, \binits{I.}}:
\batitle{Event-based vision sensor for fast and dense single-molecule
  localization microscopy}.
\bjtitle{Nature Photonics}
\bvolume{17}(\bissue{12}),
\bfpage{1105}--\blpage{1113}
(\byear{2023})
\end{barticle}
\endbibitem

%%% 21
\bibitem[\protect\citeauthoryear{Hoang}{2023}]{hoang2023neuromorphic}
\begin{botherref}
\oauthor{\bsnm{Hoang}, \binits{J.}}:
Neuromorphic cameras for atmospheric cherenkov telescopes and fast optical
  astronomy: new paradigm, challenges and opportunities.
arXiv preprint arXiv:2310.16321
(2023)
\end{botherref}
\endbibitem

%%% 22
\bibitem[\protect\citeauthoryear{Meddi et~al.}{2012}]{meddi2012new}
\begin{barticle}
\bauthor{\bsnm{Meddi}, \binits{F.}},
\bauthor{\bsnm{Ambrosino}, \binits{F.}},
\bauthor{\bsnm{Nesci}, \binits{R.}},
\bauthor{\bsnm{Rossi}, \binits{C.}},
\bauthor{\bsnm{Sclavi}, \binits{S.}},
\bauthor{\bsnm{Bruni}, \binits{I.}},
\bauthor{\bsnm{Ruggieri}, \binits{A.}},
\bauthor{\bsnm{Sestito}, \binits{S.}}:
\batitle{A new fast silicon photomultiplier photometer1}.
\bjtitle{Publications of the Astronomical Society of the Pacific}
\bvolume{124}(\bissue{915}),
\bfpage{448}
(\byear{2012})
\end{barticle}
\endbibitem

%%% 23
\bibitem[\protect\citeauthoryear{Sen et~al.}{2022}]{sen2022astronomical}
\begin{barticle}
\bauthor{\bsnm{Sen}, \binits{S.}},
\bauthor{\bsnm{Agarwal}, \binits{S.}},
\bauthor{\bsnm{Chakraborty}, \binits{P.}},
\bauthor{\bsnm{Singh}, \binits{K.P.}}:
\batitle{Astronomical big data processing using machine learning: A
  comprehensive review}.
\bjtitle{Experimental Astronomy}
\bvolume{53}(\bissue{1}),
\bfpage{1}--\blpage{43}
(\byear{2022})
\end{barticle}
\endbibitem

%%% 24
\bibitem[\protect\citeauthoryear{Zhang et~al.}{2015}]{7363840}
\begin{bchapter}
\bauthor{\bsnm{Zhang}, \binits{Z.}},
\bauthor{\bsnm{Barbary}, \binits{K.}},
\bauthor{\bsnm{Nothaft}, \binits{F.A.}},
\bauthor{\bsnm{Sparks}, \binits{E.}},
\bauthor{\bsnm{Zahn}, \binits{O.}},
\bauthor{\bsnm{Franklin}, \binits{M.J.}},
\bauthor{\bsnm{Patterson}, \binits{D.A.}},
\bauthor{\bsnm{Perlmutter}, \binits{S.}}:
\bctitle{Scientific computing meets big data technology: An astronomy use
  case}.
In: \bbtitle{2015 IEEE International Conference on Big Data (Big Data)},
pp. \bfpage{918}--\blpage{927}
(\byear{2015}).
\doiurl{10.1109/BigData.2015.7363840}
\end{bchapter}
\endbibitem

%%% 25
\bibitem[\protect\citeauthoryear{Mickaelian and
  Mikayelyan}{2019}]{mickaelian2019role}
\begin{barticle}
\bauthor{\bsnm{Mickaelian}, \binits{A.}},
\bauthor{\bsnm{Mikayelyan}, \binits{G.}}:
\batitle{The role of big data in astronomy education}.
\bjtitle{Proceedings of the International Astronomical Union}
\bvolume{15}(\bissue{S367}),
\bfpage{214}--\blpage{217}
(\byear{2019})
\end{barticle}
\endbibitem

%%% 26
\bibitem[\protect\citeauthoryear{Faaique}{2024}]{faaique2024overview}
\begin{barticle}
\bauthor{\bsnm{Faaique}, \binits{M.}}:
\batitle{Overview of big data analytics in modern astronomy}.
\bjtitle{International Journal of Mathematics, Statistics, and Computer
  Science}
\bvolume{2},
\bfpage{96}--\blpage{113}
(\byear{2024})
\end{barticle}
\endbibitem

%%% 27
\bibitem[\protect\citeauthoryear{Sachdeva et~al.}{2022}]{sachdeva2023big}
\begin{bbook}
\bauthor{\bsnm{Sachdeva}, \binits{S.}},
\bauthor{\bsnm{Watanobe}, \binits{Y.}},
\bauthor{\bsnm{Bhalla}, \binits{S.}}:
\bbtitle{Big-Data-Analytics in Astronomy, Science, and Engineering}.
\bpublisher{Springer},
\blocation{USA}
(\byear{2022})
\end{bbook}
\endbibitem

%%% 28
\bibitem[\protect\citeauthoryear{Zhou et~al.}{2024}]{zhou2024bioinspired}
\begin{botherref}
\oauthor{\bsnm{Zhou}, \binits{Y.}},
\oauthor{\bsnm{Yan}, \binits{Z.}},
\oauthor{\bsnm{Yang}, \binits{Y.}},
\oauthor{\bsnm{Wang}, \binits{Z.}},
\oauthor{\bsnm{Lu}, \binits{P.}},
\oauthor{\bsnm{Yuan}, \binits{P.F.}},
\oauthor{\bsnm{He}, \binits{B.}}:
Bioinspired sensors and applications in intelligent robots: a review.
Robotic Intelligence and Automation
(2024)
\end{botherref}
\endbibitem

%%% 29
\bibitem[\protect\citeauthoryear{Kremer et~al.}{2017}]{7887648}
\begin{barticle}
\bauthor{\bsnm{Kremer}, \binits{J.}},
\bauthor{\bsnm{Stensbo-Smidt}, \binits{K.}},
\bauthor{\bsnm{Gieseke}, \binits{F.}},
\bauthor{\bsnm{Pedersen}, \binits{K.S.}},
\bauthor{\bsnm{Igel}, \binits{C.}}:
\batitle{Big universe, big data: Machine learning and image analysis for
  astronomy}.
\bjtitle{IEEE Intelligent Systems}
\bvolume{32}(\bissue{2}),
\bfpage{16}--\blpage{22}
(\byear{2017})
\doiurl{10.1109/MIS.2017.40}
\end{barticle}
\endbibitem

%%% 30
\bibitem[\protect\citeauthoryear{Zhang and Zhao}{2015}]{zhang2015astronomy}
\begin{barticle}
\bauthor{\bsnm{Zhang}, \binits{Y.}},
\bauthor{\bsnm{Zhao}, \binits{Y.}}:
\batitle{Astronomy in the big data era}.
\bjtitle{Data Science Journal}
\bvolume{14},
\bfpage{11}--\blpage{11}
(\byear{2015})
\end{barticle}
\endbibitem

%%% 31
\bibitem[\protect\citeauthoryear{Hainaut and
  Williams}{2020}]{hainaut2020impact}
\begin{barticle}
\bauthor{\bsnm{Hainaut}, \binits{O.R.}},
\bauthor{\bsnm{Williams}, \binits{A.P.}}:
\batitle{Impact of satellite constellations on astronomical observations with
  eso telescopes in the visible and infrared domains}.
\bjtitle{Astronomy \& Astrophysics}
\bvolume{636},
\bfpage{121}
(\byear{2020})
\end{barticle}
\endbibitem

%%% 32
\bibitem[\protect\citeauthoryear{Lawler et~al.}{2021}]{lawler2021visibility}
\begin{barticle}
\bauthor{\bsnm{Lawler}, \binits{S.M.}},
\bauthor{\bsnm{Boley}, \binits{A.C.}},
\bauthor{\bsnm{Rein}, \binits{H.}}:
\batitle{Visibility predictions for near-future satellite megaconstellations:
  latitudes near 50 will experience the worst light pollution}.
\bjtitle{The Astronomical Journal}
\bvolume{163}(\bissue{1}),
\bfpage{21}
(\byear{2021})
\end{barticle}
\endbibitem

%%% 33
\bibitem[\protect\citeauthoryear{Tyson et~al.}{2020}]{tyson2020mitigation}
\begin{barticle}
\bauthor{\bsnm{Tyson}, \binits{J.A.}},
\bauthor{\bsnm{Ivezi{\'c}}, \binits{{\v{Z}}.}},
\bauthor{\bsnm{Bradshaw}, \binits{A.}},
\bauthor{\bsnm{Rawls}, \binits{M.L.}},
\bauthor{\bsnm{Xin}, \binits{B.}},
\bauthor{\bsnm{Yoachim}, \binits{P.}},
\bauthor{\bsnm{Parejko}, \binits{J.}},
\bauthor{\bsnm{Greene}, \binits{J.}},
\bauthor{\bsnm{Sholl}, \binits{M.}},
\bauthor{\bsnm{Abbott}, \binits{T.M.}}, \betal:
\batitle{Mitigation of leo satellite brightness and trail effects on the rubin
  observatory lsst}.
\bjtitle{The Astronomical Journal}
\bvolume{160}(\bissue{5}),
\bfpage{226}
(\byear{2020})
\end{barticle}
\endbibitem

%%% 34
\bibitem[\protect\citeauthoryear{Kocifaj
  et~al.}{2021}]{kocifaj2021proliferation}
\begin{barticle}
\bauthor{\bsnm{Kocifaj}, \binits{M.}},
\bauthor{\bsnm{Kundracik}, \binits{F.}},
\bauthor{\bsnm{Barentine}, \binits{J.C.}},
\bauthor{\bsnm{Bar{\'a}}, \binits{S.}}:
\batitle{The proliferation of space objects is a rapidly increasing source of
  artificial night sky brightness}.
\bjtitle{Monthly Notices of the Royal Astronomical Society: Letters}
\bvolume{504}(\bissue{1}),
\bfpage{40}--\blpage{44}
(\byear{2021})
\end{barticle}
\endbibitem

%%% 35
\bibitem[\protect\citeauthoryear{Mahowald and
  Mahowald}{1994}]{mahowald1994silicon}
\begin{botherref}
\oauthor{\bsnm{Mahowald}, \binits{M.}},
\oauthor{\bsnm{Mahowald}, \binits{M.}}:
The silicon retina.
An Analog VLSI System for Stereoscopic Vision,
4--65
(1994)
\end{botherref}
\endbibitem

%%% 36
\bibitem[\protect\citeauthoryear{Gallego et~al.}{2020}]{gallego2020event}
\begin{barticle}
\bauthor{\bsnm{Gallego}, \binits{G.}},
\bauthor{\bsnm{Delbr{\"u}ck}, \binits{T.}},
\bauthor{\bsnm{Orchard}, \binits{G.}},
\bauthor{\bsnm{Bartolozzi}, \binits{C.}},
\bauthor{\bsnm{Taba}, \binits{B.}},
\bauthor{\bsnm{Censi}, \binits{A.}},
\bauthor{\bsnm{Leutenegger}, \binits{S.}},
\bauthor{\bsnm{Davison}, \binits{A.J.}},
\bauthor{\bsnm{Conradt}, \binits{J.}},
\bauthor{\bsnm{Daniilidis}, \binits{K.}}, \betal:
\batitle{Event-based vision: A survey}.
\bjtitle{IEEE transactions on pattern analysis and machine intelligence}
\bvolume{44}(\bissue{1}),
\bfpage{154}--\blpage{180}
(\byear{2020})
\end{barticle}
\endbibitem

%%% 37
\bibitem[\protect\citeauthoryear{Posch et~al.}{2014}]{posch2014retinomorphic}
\begin{barticle}
\bauthor{\bsnm{Posch}, \binits{C.}},
\bauthor{\bsnm{Serrano-Gotarredona}, \binits{T.}},
\bauthor{\bsnm{Linares-Barranco}, \binits{B.}},
\bauthor{\bsnm{Delbruck}, \binits{T.}}:
\batitle{Retinomorphic event-based vision sensors: bioinspired cameras with
  spiking output}.
\bjtitle{Proceedings of the IEEE}
\bvolume{102}(\bissue{10}),
\bfpage{1470}--\blpage{1484}
(\byear{2014})
\end{barticle}
\endbibitem

%%% 38
\bibitem[\protect\citeauthoryear{Posch et~al.}{2007}]{4252856}
\begin{bchapter}
\bauthor{\bsnm{Posch}, \binits{C.}},
\bauthor{\bsnm{Hofstatter}, \binits{M.}},
\bauthor{\bsnm{Litzenberger}, \binits{M.}},
\bauthor{\bsnm{Matolin}, \binits{D.}},
\bauthor{\bsnm{Donath}, \binits{N.}},
\bauthor{\bsnm{Schon}, \binits{P.}},
\bauthor{\bsnm{Garn}, \binits{H.}}:
\bctitle{Wide dynamic range, high-speed machine vision with a 2×256 pixel
  temporal contrast vision sensor}.
In: \bbtitle{2007 IEEE International Symposium on Circuits and Systems
  (ISCAS)},
pp. \bfpage{1196}--\blpage{1199}
(\byear{2007}).
\doiurl{10.1109/ISCAS.2007.378266}
\end{bchapter}
\endbibitem

%%% 39
\bibitem[\protect\citeauthoryear{Han et~al.}{2020}]{9156346}
\begin{bchapter}
\bauthor{\bsnm{Han}, \binits{J.}},
\bauthor{\bsnm{Zhou}, \binits{C.}},
\bauthor{\bsnm{Duan}, \binits{P.}},
\bauthor{\bsnm{Tang}, \binits{Y.}},
\bauthor{\bsnm{Xu}, \binits{C.}},
\bauthor{\bsnm{Xu}, \binits{C.}},
\bauthor{\bsnm{Huang}, \binits{T.}},
\bauthor{\bsnm{Shi}, \binits{B.}}:
\bctitle{Neuromorphic camera guided high dynamic range imaging}.
In: \bbtitle{2020 IEEE/CVF Conference on Computer Vision and Pattern
  Recognition (CVPR)},
pp. \bfpage{1727}--\blpage{1736}
(\byear{2020}).
\doiurl{10.1109/CVPR42600.2020.00180}
\end{bchapter}
\endbibitem

%%% 40
\bibitem[\protect\citeauthoryear{Messikommer
  et~al.}{2022}]{messikommer2022multi}
\begin{bchapter}
\bauthor{\bsnm{Messikommer}, \binits{N.}},
\bauthor{\bsnm{Georgoulis}, \binits{S.}},
\bauthor{\bsnm{Gehrig}, \binits{D.}},
\bauthor{\bsnm{Tulyakov}, \binits{S.}},
\bauthor{\bsnm{Erbach}, \binits{J.}},
\bauthor{\bsnm{Bochicchio}, \binits{A.}},
\bauthor{\bsnm{Li}, \binits{Y.}},
\bauthor{\bsnm{Scaramuzza}, \binits{D.}}:
\bctitle{Multi-bracket high dynamic range imaging with event cameras}.
In: \bbtitle{Proceedings of the IEEE/CVF Conference on Computer Vision and
  Pattern Recognition},
pp. \bfpage{547}--\blpage{557}
(\byear{2022})
\end{bchapter}
\endbibitem

%%% 41
\bibitem[\protect\citeauthoryear{Censi and Scaramuzza}{2014}]{6906931}
\begin{bchapter}
\bauthor{\bsnm{Censi}, \binits{A.}},
\bauthor{\bsnm{Scaramuzza}, \binits{D.}}:
\bctitle{Low-latency event-based visual odometry}.
In: \bbtitle{2014 IEEE International Conference on Robotics and Automation
  (ICRA)},
pp. \bfpage{703}--\blpage{710}
(\byear{2014}).
\doiurl{10.1109/ICRA.2014.6906931}
\end{bchapter}
\endbibitem

%%% 42
\bibitem[\protect\citeauthoryear{Delbruck et~al.}{2008}]{delbruck2008frame}
\begin{bchapter}
\bauthor{\bsnm{Delbruck}, \binits{T.}}, \betal:
\bctitle{Frame-free dynamic digital vision}.
In: \bbtitle{Proceedings of Intl. Symp. on Secure-Life Electronics, Advanced
  Electronics for Quality Life and Society},
vol. \bseriesno{1},
pp. \bfpage{21}--\blpage{26}
(\byear{2008})
\end{bchapter}
\endbibitem

%%% 43
\bibitem[\protect\citeauthoryear{Leñero-Bardallo et~al.}{2011}]{lenero20113}
\begin{barticle}
\bauthor{\bsnm{Leñero-Bardallo}, \binits{J.A.}},
\bauthor{\bsnm{Serrano-Gotarredona}, \binits{T.}},
\bauthor{\bsnm{Linares-Barranco}, \binits{B.}}:
\batitle{A 3.6 $\mu$ s latency asynchronous frame-free event-driven
  dynamic-vision-sensor}.
\bjtitle{IEEE Journal of Solid-State Circuits}
\bvolume{46}(\bissue{6}),
\bfpage{1443}--\blpage{1455}
(\byear{2011})
\doiurl{10.1109/JSSC.2011.2118490}
\end{barticle}
\endbibitem

%%% 44
\bibitem[\protect\citeauthoryear{Serrano-Gotarredona and
  Linares-Barranco}{2013}]{6407468}
\begin{barticle}
\bauthor{\bsnm{Serrano-Gotarredona}, \binits{T.}},
\bauthor{\bsnm{Linares-Barranco}, \binits{B.}}:
\batitle{A 128 $\times$ 128 1.5$\%$ contrast sensitivity 0.9$\%$ fpn 3 $\mu$s
  latency 4 mw asynchronous frame-free dynamic vision sensor using
  transimpedance preamplifiers}.
\bjtitle{IEEE Journal of Solid-State Circuits}
\bvolume{48}(\bissue{3}),
\bfpage{827}--\blpage{838}
(\byear{2013})
\doiurl{10.1109/JSSC.2012.2230553}
\end{barticle}
\endbibitem

%%% 45
\bibitem[\protect\citeauthoryear{Cohen et~al.}{2019}]{cohen2019event}
\begin{barticle}
\bauthor{\bsnm{Cohen}, \binits{G.}},
\bauthor{\bsnm{Afshar}, \binits{S.}},
\bauthor{\bsnm{Morreale}, \binits{B.}},
\bauthor{\bsnm{Bessell}, \binits{T.}},
\bauthor{\bsnm{Wabnitz}, \binits{A.}},
\bauthor{\bsnm{Rutten}, \binits{M.}},
\bauthor{\bsnm{Schaik}, \binits{A.}}:
\batitle{Event-based sensing for space situational awareness}.
\bjtitle{The Journal of the Astronautical Sciences}
\bvolume{66},
\bfpage{125}--\blpage{141}
(\byear{2019})
\end{barticle}
\endbibitem

%%% 46
\bibitem[\protect\citeauthoryear{Afshar et~al.}{2020}]{9142352}
\begin{barticle}
\bauthor{\bsnm{Afshar}, \binits{S.}},
\bauthor{\bsnm{Nicholson}, \binits{A.P.}},
\bauthor{\bsnm{Schaik}, \binits{A.}},
\bauthor{\bsnm{Cohen}, \binits{G.}}:
\batitle{Event-based object detection and tracking for space situational
  awareness}.
\bjtitle{IEEE Sensors Journal}
\bvolume{20}(\bissue{24}),
\bfpage{15117}--\blpage{15132}
(\byear{2020})
\doiurl{10.1109/JSEN.2020.3009687}
\end{barticle}
\endbibitem

%%% 47
\bibitem[\protect\citeauthoryear{McMahon-Crabtree and
  Monet}{2021}]{mcmahon2021commercial}
\begin{barticle}
\bauthor{\bsnm{McMahon-Crabtree}, \binits{P.N.}},
\bauthor{\bsnm{Monet}, \binits{D.G.}}:
\batitle{Commercial-off-the-shelf event-based cameras for space surveillance
  applications}.
\bjtitle{Applied Optics}
\bvolume{60}(\bissue{25}),
\bfpage{144}--\blpage{153}
(\byear{2021})
\end{barticle}
\endbibitem

%%% 48
\bibitem[\protect\citeauthoryear{Boehrer et~al.}{2019}]{boehrer2019using}
\begin{bchapter}
\bauthor{\bsnm{Boehrer}, \binits{N.}},
\bauthor{\bsnm{Nieuwenhuizen}, \binits{R.}},
\bauthor{\bsnm{Dijk}, \binits{J.}}:
\bctitle{Using event cameras for imaging through atmospheric turbulence}.
In: \bbtitle{COAT-2019-workshop (Communications and Observations Through
  Atmospheric Turbulence: Characterization and Mitigation)}
(\byear{2019})
\end{bchapter}
\endbibitem

%%% 49
\bibitem[\protect\citeauthoryear{Polnau and
  Vorontsov}{2021}]{polnau2021atmospheric}
\begin{barticle}
\bauthor{\bsnm{Polnau}, \binits{E.}},
\bauthor{\bsnm{Vorontsov}, \binits{M.A.}}:
\batitle{Atmospheric turbulence characterization using a neuromorphic
  camera-based imaging sensor}.
\bjtitle{Journal of Optics}
\bvolume{23}(\bissue{12}),
\bfpage{125608}
(\byear{2021})
\end{barticle}
\endbibitem

%%% 50
\bibitem[\protect\citeauthoryear{Cohen et~al.}{2022}]{cohen2022exploring}
\begin{botherref}
\oauthor{\bsnm{Cohen}, \binits{G.}},
\oauthor{\bsnm{Kong}, \binits{F.}},
\oauthor{\bsnm{Lambert}, \binits{A.}},
\oauthor{\bsnm{UNIVERSITY}, \binits{W.S.}}:
Exploring the use of event based imaging sensors on turbulence characterisation
  and adaptive optics
(2022)
\end{botherref}
\endbibitem

%%% 51
\bibitem[\protect\citeauthoryear{Kong et~al.}{2019}]{kong2019using}
\begin{bchapter}
\bauthor{\bsnm{Kong}, \binits{F.}},
\bauthor{\bsnm{Cohen}, \binits{G.}},
\bauthor{\bsnm{Lambert}, \binits{A.}}:
\bctitle{Using event-based optical flow to determine the shack-hartmann spot
  displacements}.
In: \bbtitle{Propagation Through and Characterization of Atmospheric and
  Oceanic Phenomena},
pp. \bfpage{4}--\blpage{3}
(\byear{2019}).
\bcomment{Optica Publishing Group}
\end{bchapter}
\endbibitem

%%% 52
\bibitem[\protect\citeauthoryear{Kong et~al.}{2020}]{kong2020shack}
\begin{barticle}
\bauthor{\bsnm{Kong}, \binits{F.}},
\bauthor{\bsnm{Lambert}, \binits{A.}},
\bauthor{\bsnm{Joubert}, \binits{D.}},
\bauthor{\bsnm{Cohen}, \binits{G.}}:
\batitle{Shack-hartmann wavefront sensing using spatial-temporal data from an
  event-based image sensor}.
\bjtitle{Optics Express}
\bvolume{28}(\bissue{24}),
\bfpage{36159}--\blpage{36175}
(\byear{2020})
\end{barticle}
\endbibitem

%%% 53
\bibitem[\protect\citeauthoryear{Wisentaner}{2022}]{wisentaner2022orbit}
\begin{botherref}
\oauthor{\bsnm{Wisentaner}, \binits{C.M.}}:
Orbit determination with event-based cameras to improve space domain awareness
(2022).
Air Force Institute of Technology, Theses and Dissertations. Available at:
  \url{https://scholar.afit.edu/etd/5544}
\end{botherref}
\endbibitem

%%% 54
\bibitem[\protect\citeauthoryear{Jolley et~al.}{2022}]{jolley2022evaluation}
\begin{barticle}
\bauthor{\bsnm{Jolley}, \binits{A.}},
\bauthor{\bsnm{Cohen}, \binits{G.}},
\bauthor{\bsnm{Joubert}, \binits{D.}},
\bauthor{\bsnm{Lambert}, \binits{A.}}:
\batitle{Evaluation of event-based sensors for satellite material
  characterization}.
\bjtitle{Journal of Spacecraft and Rockets}
\bvolume{59}(\bissue{2}),
\bfpage{627}--\blpage{636}
(\byear{2022})
\end{barticle}
\endbibitem

%%% 55
\bibitem[\protect\citeauthoryear{Jolley et~al.}{2023}]{jolley2023neuromorphic}
\begin{barticle}
\bauthor{\bsnm{Jolley}, \binits{A.}},
\bauthor{\bsnm{Afshar}, \binits{S.}},
\bauthor{\bsnm{Cohen}, \binits{G.}},
\bauthor{\bsnm{Lazarus~Pahlavani}, \binits{R.}},
\bauthor{\bsnm{Lambert}, \binits{A.}}:
\batitle{Neuromorphic sensor event-rate monitoring for satellite
  characterization}.
\bjtitle{Journal of Spacecraft and Rockets}
\bvolume{60}(\bissue{3}),
\bfpage{753}--\blpage{764}
(\byear{2023})
\end{barticle}
\endbibitem

%%% 56
\bibitem[\protect\citeauthoryear{Jolley and
  Cohen}{2022}]{jolley2022characterising}
\begin{barticle}
\bauthor{\bsnm{Jolley}, \binits{A.}},
\bauthor{\bsnm{Cohen}, \binits{G.}}:
\batitle{Characterising satellites using neuromorphic sensor multicolour
  broadband event-rates}.
\bjtitle{44th COSPAR Scientific Assembly. Held 16-24 July}
\bvolume{44},
\bfpage{3162}
(\byear{2022})
\end{barticle}
\endbibitem

%%% 57
\bibitem[\protect\citeauthoryear{Sikorski et~al.}{2021}]{sikorski2021event}
\begin{bchapter}
\bauthor{\bsnm{Sikorski}, \binits{O.}},
\bauthor{\bsnm{Izzo}, \binits{D.}},
\bauthor{\bsnm{Meoni}, \binits{G.}}:
\bctitle{Event-based spacecraft landing using time-to-contact}.
In: \bbtitle{Proceedings of the IEEE/CVF Conference on Computer Vision and
  Pattern Recognition},
pp. \bfpage{1941}--\blpage{1950}
(\byear{2021})
\end{bchapter}
\endbibitem

%%% 58
\bibitem[\protect\citeauthoryear{Ralph et~al.}{2022}]{ralph2022real}
\begin{barticle}
\bauthor{\bsnm{Ralph}, \binits{N.}},
\bauthor{\bsnm{Joubert}, \binits{D.}},
\bauthor{\bsnm{Jolley}, \binits{A.}},
\bauthor{\bsnm{Afshar}, \binits{S.}},
\bauthor{\bsnm{Tothill}, \binits{N.}},
\bauthor{\bsnm{Van~Schaik}, \binits{A.}},
\bauthor{\bsnm{Cohen}, \binits{G.}}:
\batitle{Real-time event-based unsupervised feature consolidation and tracking
  for space situational awareness}.
\bjtitle{Frontiers in neuroscience}
\bvolume{16},
\bfpage{821157}
(\byear{2022})
\end{barticle}
\endbibitem

%%% 59
\bibitem[\protect\citeauthoryear{Felsen et~al.}{2022}]{felsen_detecting_2022}
\begin{bchapter}
\bauthor{\bsnm{Felsen}, \binits{P.}},
\bauthor{\bsnm{Scrofano}, \binits{R.}},
\bauthor{\bsnm{Rosales}, \binits{R.}},
\bauthor{\bsnm{Subasavage}, \binits{J.}},
\bauthor{\bsnm{Desai}, \binits{N.}},
\bauthor{\bsnm{Smith}, \binits{T.}},
\bauthor{\bsnm{Dearborn}, \binits{M.}}:
\bctitle{Detecting space objects in event camera data through 3d point cloud
  processing}.
In: \bbtitle{Proceedings of the Advanced Maui Optical and Space Surveillance
  Technologies Conference}
(\byear{2022})
\end{bchapter}
\endbibitem

%%% 60
\bibitem[\protect\citeauthoryear{Ralph et~al.}{2023}]{ralph2023shake}
\begin{bchapter}
\bauthor{\bsnm{Ralph}, \binits{N.}},
\bauthor{\bsnm{Maybour}, \binits{D.}},
\bauthor{\bsnm{Marcireau}, \binits{A.}},
\bauthor{\bsnm{Jones}, \binits{I.}},
\bauthor{\bsnm{De~Horta}, \binits{A.}},
\bauthor{\bsnm{Cohen}, \binits{G.}}:
\bctitle{Shake before use: Artificial contrast generation for improved space
  imaging using neuromorphic event-based vision sensors}.
In: \bbtitle{Proceedings of the Advanced Maui Optical and Space Surveillance
  (AMOS) Technologies Conference},
p. \bfpage{161}
(\byear{2023})
\end{bchapter}
\endbibitem

%%% 61
\bibitem[\protect\citeauthoryear{Afshar et~al.}{2019}]{afshar2019investigation}
\begin{barticle}
\bauthor{\bsnm{Afshar}, \binits{S.}},
\bauthor{\bsnm{Hamilton}, \binits{T.J.}},
\bauthor{\bsnm{Tapson}, \binits{J.}},
\bauthor{\bsnm{Van~Schaik}, \binits{A.}},
\bauthor{\bsnm{Cohen}, \binits{G.}}:
\batitle{Investigation of event-based surfaces for high-speed detection,
  unsupervised feature extraction, and object recognition}.
\bjtitle{Frontiers in neuroscience}
\bvolume{12},
\bfpage{1047}
(\byear{2019})
\end{barticle}
\endbibitem

%%% 62
\bibitem[\protect\citeauthoryear{Dong et~al.}{2023}]{dong2023event}
\begin{bchapter}
\bauthor{\bsnm{Dong}, \binits{P.}},
\bauthor{\bsnm{Yue}, \binits{M.}},
\bauthor{\bsnm{Zhu}, \binits{L.}},
\bauthor{\bsnm{Xu}, \binits{F.}},
\bauthor{\bsnm{Du}, \binits{Z.}}:
\bctitle{Event-based weak target detection and tracking for space situational
  awareness}.
In: \bbtitle{2023 International Conference on Cyber-Enabled Distributed
  Computing and Knowledge Discovery (CyberC)},
pp. \bfpage{78}--\blpage{83}
(\byear{2023}).
\bcomment{IEEE}
\end{bchapter}
\endbibitem

%%% 63
\bibitem[\protect\citeauthoryear{Westerhout
  et~al.}{2023}]{westerhout2023analysis}
\begin{bchapter}
\bauthor{\bsnm{Westerhout}, \binits{V.}},
\bauthor{\bsnm{Valdivia}, \binits{S.}},
\bauthor{\bsnm{Vera}, \binits{E.}}:
\bctitle{Analysis of detection limits in event-based cameras for space
  situational awareness}.
In: \bbtitle{Proceedings of the Advanced Maui Optical and Space Surveillance
  (AMOS) Technologies Conference},
p. \bfpage{197}
(\byear{2023})
\end{bchapter}
\endbibitem

%%% 64
\bibitem[\protect\citeauthoryear{{\.Z}o{\l}nowski
  et~al.}{2019}]{zolnowski2019observational}
\begin{bchapter}
\bauthor{\bsnm{{\.Z}o{\l}nowski}, \binits{M.}},
\bauthor{\bsnm{Reszelewski}, \binits{R.}},
\bauthor{\bsnm{Moeys}, \binits{D.P.}},
\bauthor{\bsnm{Delbr{\"u}ck}, \binits{T.}},
\bauthor{\bsnm{Kami{\'n}ski}, \binits{K.}}:
\bctitle{Observational evaluation of event cameras performance in optical space
  surveillance}.
In: \bbtitle{NEO and Debris Detection Conference, Darmstadt, Germany}
(\byear{2019})
\end{bchapter}
\endbibitem

%%% 65
\bibitem[\protect\citeauthoryear{Cohen et~al.}{2018}]{cohen2018approaches}
\begin{bchapter}
\bauthor{\bsnm{Cohen}, \binits{G.}},
\bauthor{\bsnm{Afshar}, \binits{S.}},
\bauthor{\bsnm{Van~Schaik}, \binits{A.}}:
\bctitle{Approaches for astrometry using event-based sensors}.
In: \bbtitle{Advanced Maui Optical and Space Surveillance (AMOS) Technologies
  Conference},
p. \bfpage{25}
(\byear{2018})
\end{bchapter}
\endbibitem

%%% 66
\bibitem[\protect\citeauthoryear{Ralph et~al.}{2023}]{ralph2023astrometric}
\begin{barticle}
\bauthor{\bsnm{Ralph}, \binits{N.O.}},
\bauthor{\bsnm{Marcireau}, \binits{A.}},
\bauthor{\bsnm{Afshar}, \binits{S.}},
\bauthor{\bsnm{Tothill}, \binits{N.}},
\bauthor{\bsnm{Van~Schaik}, \binits{A.}},
\bauthor{\bsnm{Cohen}, \binits{G.}}:
\batitle{Astrometric calibration and source characterisation of the latest
  generation neuromorphic event-based cameras for space imaging}.
\bjtitle{Astrodynamics}
\bvolume{7}(\bissue{4}),
\bfpage{415}--\blpage{443}
(\byear{2023})
\end{barticle}
\endbibitem

%%% 67
\bibitem[\protect\citeauthoryear{Chin et~al.}{2019}]{chin2019star}
\begin{bchapter}
\bauthor{\bsnm{Chin}, \binits{T.-J.}},
\bauthor{\bsnm{Bagchi}, \binits{S.}},
\bauthor{\bsnm{Eriksson}, \binits{A.}},
\bauthor{\bsnm{Van~Schaik}, \binits{A.}}:
\bctitle{Star tracking using an event camera}.
In: \bbtitle{Proceedings of the IEEE/CVF Conference on Computer Vision and
  Pattern Recognition Workshops},
pp. \bfpage{0}--\blpage{0}
(\byear{2019})
\end{bchapter}
\endbibitem

%%% 68
\bibitem[\protect\citeauthoryear{McHarg et~al.}{2022}]{mcharg2022falcon}
\begin{barticle}
\bauthor{\bsnm{McHarg}, \binits{M.G.}},
\bauthor{\bsnm{Balthazor}, \binits{R.L.}},
\bauthor{\bsnm{McReynolds}, \binits{B.J.}},
\bauthor{\bsnm{Howe}, \binits{D.H.}},
\bauthor{\bsnm{Maloney}, \binits{C.J.}},
\bauthor{\bsnm{O’Keefe}, \binits{D.}},
\bauthor{\bsnm{Bam}, \binits{R.}},
\bauthor{\bsnm{Wilson}, \binits{G.}},
\bauthor{\bsnm{Karki}, \binits{P.}},
\bauthor{\bsnm{Marcireau}, \binits{A.}}, \betal:
\batitle{Falcon neuro: an event-based sensor on the international space
  station}.
\bjtitle{Optical Engineering}
\bvolume{61}(\bissue{8}),
\bfpage{085105}--\blpage{085105}
(\byear{2022})
\end{barticle}
\endbibitem

%%% 69
\bibitem[\protect\citeauthoryear{McHarg et~al.}{2020}]{mcharg2020falcon}
\begin{bchapter}
\bauthor{\bsnm{McHarg}, \binits{M.G.}},
\bauthor{\bsnm{Harley}, \binits{J.}},
\bauthor{\bsnm{Balthazor}, \binits{R.L.}},
\bauthor{\bsnm{Wilson}, \binits{G.}},
\bauthor{\bsnm{McReynolds}, \binits{B.J.}},
\bauthor{\bsnm{Cohen}, \binits{G.}},
\bauthor{\bsnm{Howe}, \binits{D.}},
\bauthor{\bsnm{Colin}, \binits{M.}}:
\bctitle{Falcon neuro—neuromorphic cameras for sprite and lightning detection
  on the international space station}.
In: \bbtitle{AGU Fall Meeting Abstracts},
vol. \bseriesno{2020},
pp. \bfpage{012}--\blpage{0002}
(\byear{2020})
\end{bchapter}
\endbibitem

%%% 70
\bibitem[\protect\citeauthoryear{Lichtsteiner}{2003}]{lichtsteiner200364x64}
\begin{bchapter}
\bauthor{\bsnm{Lichtsteiner}, \binits{P.}}:
\bctitle{64x64 event-driven logarithmic temporal derivative silicon retina}.
In: \bbtitle{Program 2003 IEEE Workshop on CCD and AIS}
(\byear{2003})
\end{bchapter}
\endbibitem

%%% 71
\bibitem[\protect\citeauthoryear{Lichtsteiner and
  Delbruck}{2005}]{lichtsteiner200564x64}
\begin{bchapter}
\bauthor{\bsnm{Lichtsteiner}, \binits{P.}},
\bauthor{\bsnm{Delbruck}, \binits{T.}}:
\bctitle{A 64x64 aer logarithmic temporal derivative silicon retina}.
In: \bbtitle{Research in Microelectronics and Electronics, 2005 PhD},
vol. \bseriesno{2},
pp. \bfpage{202}--\blpage{205}
(\byear{2005}).
\bcomment{IEEE}
\end{bchapter}
\endbibitem

%%% 72
\bibitem[\protect\citeauthoryear{Lichtsteiner
  et~al.}{2006}]{lichtsteiner2006128}
\begin{bchapter}
\bauthor{\bsnm{Lichtsteiner}, \binits{P.}},
\bauthor{\bsnm{Posch}, \binits{C.}},
\bauthor{\bsnm{Delbruck}, \binits{T.}}:
\bctitle{A 128 x 128 120db 30mw asynchronous vision sensor that responds to
  relative intensity change}.
In: \bbtitle{2006 IEEE International Solid State Circuits Conference-Digest of
  Technical Papers},
pp. \bfpage{2060}--\blpage{2069}
(\byear{2006}).
\bcomment{IEEE}
\end{bchapter}
\endbibitem

%%% 73
\bibitem[\protect\citeauthoryear{Boahen}{2004}]{boahen2004burst}
\begin{barticle}
\bauthor{\bsnm{Boahen}, \binits{K.A.}}:
\batitle{A burst-mode word-serial address-event link-i: Transmitter design}.
\bjtitle{IEEE Transactions on Circuits and Systems I: Regular Papers}
\bvolume{51}(\bissue{7}),
\bfpage{1269}--\blpage{1280}
(\byear{2004})
\end{barticle}
\endbibitem

%%% 74
\bibitem[\protect\citeauthoryear{Liu et~al.}{2014}]{liu2014event}
\begin{bbook}
\bauthor{\bsnm{Liu}, \binits{S.-C.}},
\bauthor{\bsnm{Delbruck}, \binits{T.}},
\bauthor{\bsnm{Indiveri}, \binits{G.}},
\bauthor{\bsnm{Whatley}, \binits{A.}},
\bauthor{\bsnm{Douglas}, \binits{R.}}:
\bbtitle{Event-based Neuromorphic Systems}.
\bpublisher{John Wiley \& Sons},
\blocation{Hoboken, NJ, USA}
(\byear{2014})
\end{bbook}
\endbibitem

%%% 75
\bibitem[\protect\citeauthoryear{Skorka and Joseph}{2014}]{skorka2014cmos}
\begin{bchapter}
\bauthor{\bsnm{Skorka}, \binits{O.}},
\bauthor{\bsnm{Joseph}, \binits{D.}}:
\bctitle{Cmos digital pixel sensors: technology and applications}.
In: \bbtitle{Nanosensors, Biosensors, and Info-Tech Sensors and Systems 2014},
vol. \bseriesno{9060},
pp. \bfpage{79}--\blpage{92}
(\byear{2014}).
\bcomment{SPIE}
\end{bchapter}
\endbibitem

%%% 76
\bibitem[\protect\citeauthoryear{Lichtensteiner
  et~al.}{2008}]{lichtensteiner2008128x128}
\begin{botherref}
\oauthor{\bsnm{Lichtensteiner}, \binits{P.}},
\oauthor{\bsnm{Posch}, \binits{C.}},
\oauthor{\bsnm{Delbruck}, \binits{T.}}:
A 128x128 120db 15$\mu$s latency asynchronous temporal contrast vision sensor.
IEEE Journal of Solid-State Circuits
(2),
566--576
(2008)
\end{botherref}
\endbibitem

%%% 77
\bibitem[\protect\citeauthoryear{Rebecq et~al.}{2021}]{8946715}
\begin{barticle}
\bauthor{\bsnm{Rebecq}, \binits{H.}},
\bauthor{\bsnm{Ranftl}, \binits{R.}},
\bauthor{\bsnm{Koltun}, \binits{V.}},
\bauthor{\bsnm{Scaramuzza}, \binits{D.}}:
\batitle{High speed and high dynamic range video with an event camera}.
\bjtitle{IEEE Transactions on Pattern Analysis and Machine Intelligence}
\bvolume{43}(\bissue{6}),
\bfpage{1964}--\blpage{1980}
(\byear{2021})
\doiurl{10.1109/TPAMI.2019.2963386}
\end{barticle}
\endbibitem

%%% 78
\bibitem[\protect\citeauthoryear{Scheerlinck
  et~al.}{2020}]{Scheerlinck_2020_WACV}
\begin{bchapter}
\bauthor{\bsnm{Scheerlinck}, \binits{C.}},
\bauthor{\bsnm{Rebecq}, \binits{H.}},
\bauthor{\bsnm{Gehrig}, \binits{D.}},
\bauthor{\bsnm{Barnes}, \binits{N.}},
\bauthor{\bsnm{Mahony}, \binits{R.}},
\bauthor{\bsnm{Scaramuzza}, \binits{D.}}:
\bctitle{Fast image reconstruction with an event camera}.
In: \bbtitle{Proceedings of the IEEE/CVF Winter Conference on Applications of
  Computer Vision (WACV)}
(\byear{2020})
\end{bchapter}
\endbibitem

%%% 79
\bibitem[\protect\citeauthoryear{Sharma et~al.}{2019}]{8702508}
\begin{bchapter}
\bauthor{\bsnm{Sharma}, \binits{R.}},
\bauthor{\bsnm{Gupta}, \binits{S.}},
\bauthor{\bsnm{Kumar}, \binits{K.}},
\bauthor{\bsnm{Kumar}, \binits{P.}},
\bauthor{\bsnm{Thakur}, \binits{C.S.}}:
\bctitle{Real-time image segmentation using neuromorphic pixel array}.
In: \bbtitle{2019 IEEE International Symposium on Circuits and Systems
  (ISCAS)},
pp. \bfpage{1}--\blpage{5}
(\byear{2019}).
\doiurl{10.1109/ISCAS.2019.8702508}
\end{bchapter}
\endbibitem

%%% 80
\bibitem[\protect\citeauthoryear{Finateu et~al.}{2020}]{9063149}
\begin{bchapter}
\bauthor{\bsnm{Finateu}, \binits{T.}},
\bauthor{\bsnm{Niwa}, \binits{A.}},
\bauthor{\bsnm{Matolin}, \binits{D.}},
\bauthor{\bsnm{Tsuchimoto}, \binits{K.}},
\bauthor{\bsnm{Mascheroni}, \binits{A.}},
\bauthor{\bsnm{Reynaud}, \binits{E.}},
\bauthor{\bsnm{Mostafalu}, \binits{P.}},
\bauthor{\bsnm{Brady}, \binits{F.}},
\bauthor{\bsnm{Chotard}, \binits{L.}},
\bauthor{\bsnm{LeGoff}, \binits{F.}},
\bauthor{\bsnm{Takahashi}, \binits{H.}},
\bauthor{\bsnm{Wakabayashi}, \binits{H.}},
\bauthor{\bsnm{Oike}, \binits{Y.}},
\bauthor{\bsnm{Posch}, \binits{C.}}:
\bctitle{5.10 a 1280×720 back-illuminated stacked temporal contrast
  event-based vision sensor with 4.86µm pixels, 1.066geps readout,
  programmable event-rate controller and compressive data-formatting pipeline}.
In: \bbtitle{2020 IEEE International Solid-State Circuits Conference -
  (ISSCC)},
pp. \bfpage{112}--\blpage{114}
(\byear{2020}).
\doiurl{10.1109/ISSCC19947.2020.9063149}
\end{bchapter}
\endbibitem

%%% 81
\bibitem[\protect\citeauthoryear{Lv et~al.}{2024}]{lv2024denoising}
\begin{bchapter}
\bauthor{\bsnm{Lv}, \binits{Y.}},
\bauthor{\bsnm{Liu}, \binits{Z.}},
\bauthor{\bsnm{Zhou}, \binits{L.}},
\bauthor{\bsnm{Qiao}, \binits{W.}},
\bauthor{\bsnm{Zhang}, \binits{H.}}:
\bctitle{Denoising algorithm based on event camera}.
In: \bbtitle{Sixth Conference on Frontiers in Optical Imaging and Technology:
  Novel Detector Technologies},
vol. \bseriesno{13154},
pp. \bfpage{59}--\blpage{65}
(\byear{2024}).
\bcomment{SPIE}
\end{bchapter}
\endbibitem

%%% 82
\bibitem[\protect\citeauthoryear{Annamalai and
  Thakur}{2024}]{annamalai2024beyond}
\begin{barticle}
\bauthor{\bsnm{Annamalai}, \binits{L.}},
\bauthor{\bsnm{Thakur}, \binits{C.S.}}:
\batitle{Beyond supervision: An unsupervised spatio-temporal point cloud noise
  modeling for event vision sensor}.
\bjtitle{Pattern Recognition Letters}
\bvolume{184},
\bfpage{162}--\blpage{168}
(\byear{2024})
\end{barticle}
\endbibitem

%%% 83
\bibitem[\protect\citeauthoryear{Joshi et~al.}{2022}]{joshi2022aries}
\begin{barticle}
\bauthor{\bsnm{Joshi}, \binits{Y.}},
\bauthor{\bsnm{Bangia}, \binits{T.}},
\bauthor{\bsnm{Jaiswar}, \binits{M.}},
\bauthor{\bsnm{Pant}, \binits{J.}},
\bauthor{\bsnm{Reddy}, \binits{K.}},
\bauthor{\bsnm{Yadav}, \binits{S.}}:
\batitle{Aries 130-cm devasthal fast optical telescope—operation and
  outcome}.
\bjtitle{Journal of Astronomical Instrumentation}
\bvolume{11}(\bissue{04}),
\bfpage{2240004}
(\byear{2022})
\end{barticle}
\endbibitem

%%% 84
\bibitem[\protect\citeauthoryear{Price-Whelan et~al.}{2022}]{astropy:2022}
\begin{barticle}
\bauthor{\bsnm{Price-Whelan}, \binits{A.M.}},
\bauthor{\bsnm{Lim}, \binits{P.L.}},
\bauthor{\bsnm{Earl}, \binits{N.}},
\bauthor{\bsnm{Starkman}, \binits{N.}},
\bauthor{\bsnm{Bradley}, \binits{L.}},
\bauthor{\bsnm{Shupe}, \binits{D.L.}},
\bauthor{\bsnm{Patil}, \binits{A.A.}},
\bauthor{\bsnm{Corrales}, \binits{L.}},
\bauthor{\bsnm{Brasseur}, \binits{C.}},
\bauthor{\bsnm{N{\"o}the}, \binits{M.}}, \betal:
\batitle{The astropy project: sustaining and growing a community-oriented
  open-source project and the latest major release (v5. 0) of the core
  package}.
\bjtitle{The Astrophysical Journal}
\bvolume{935}(\bissue{2}),
\bfpage{167}
(\byear{2022})
\end{barticle}
\endbibitem

%%% 85
\bibitem[\protect\citeauthoryear{Collaboration
  et~al.}{2021}]{gaia_collaboration_gaia_2021}
\begin{botherref}
\oauthor{\bsnm{Collaboration}, \binits{G.}},
\oauthor{\bsnm{Brown}, \binits{A.G.A.}},
\oauthor{\bsnm{Vallenari}, \binits{A.}},
\oauthor{\bsnm{Prusti}, \binits{T.}},
\oauthor{\bsnm{{de Bruijne}}, \binits{J.H.J.}},
\oauthor{\bsnm{Babusiaux}, \binits{C.}},
\oauthor{\bsnm{Biermann}, \binits{M.}},
\oauthor{\bsnm{Evans}, \binits{D.W.}}, et al.:
{Gaia} {Early} {Data} {Release} 3: {Summary} of the contents and survey
  properties.
arXiv
(2021)
{\href{https://arxiv.org/abs/2012.01533}{{arXiv:2012.01533}}}
{[astro-ph.IM]}.
Accessed 2021-06-03
\end{botherref}
\endbibitem

%%% 86
\bibitem[\protect\citeauthoryear{O'Dell and Wong}{1995}]{odell1995}
\begin{botherref}
\oauthor{\bsnm{O'Dell}, \binits{C.R.}},
\oauthor{\bsnm{Wong}, \binits{S.K.}}:
STScI-PRC1995-45a.
\url{http://hubblesite.org}.
Credit: NASA, C.R. O'Dell and S.K. Wong (Rice University)
(1995)
\end{botherref}
\endbibitem

%%% 87
\bibitem[\protect\citeauthoryear{{NASA}}{}]{nasa_lunar_impact_2024}
\begin{botherref}
\oauthor{\bsnm{{NASA}}}:
Lunar Impact Monitoring.
\url{https://www.nasa.gov/meteoroid-environment-office/about-lunar-impact-monitoring/}.
Accessed: 19 August 2024
\end{botherref}
\endbibitem

%%% 88
\bibitem[\protect\citeauthoryear{Cudnik et~al.}{2003}]{cudnik2003observation}
\begin{barticle}
\bauthor{\bsnm{Cudnik}, \binits{B.M.}},
\bauthor{\bsnm{Palmer}, \binits{D.W.}},
\bauthor{\bsnm{Palmer}, \binits{D.M.}},
\bauthor{\bsnm{Cook}, \binits{A.}},
\bauthor{\bsnm{Venable}, \binits{R.}},
\bauthor{\bsnm{Gural}, \binits{P.S.}}:
\batitle{The observation and characterization of lunar meteoroid impact
  phenomena}.
\bjtitle{Earth, Moon, and Planets}
\bvolume{93},
\bfpage{97}--\blpage{106}
(\byear{2003})
\end{barticle}
\endbibitem

%%% 89
\bibitem[\protect\citeauthoryear{Pokorn{\`y}
  et~al.}{2019}]{pokorny2019meteoroids}
\begin{barticle}
\bauthor{\bsnm{Pokorn{\`y}}, \binits{P.}},
\bauthor{\bsnm{Janches}, \binits{D.}},
\bauthor{\bsnm{Sarantos}, \binits{M.}},
\bauthor{\bsnm{Szalay}, \binits{J.R.}},
\bauthor{\bsnm{Hor{\'a}nyi}, \binits{M.}},
\bauthor{\bsnm{Nesvorn{\`y}}, \binits{D.}},
\bauthor{\bsnm{Kuchner}, \binits{M.J.}}:
\batitle{Meteoroids at the moon: orbital properties, surface vaporization, and
  impact ejecta production}.
\bjtitle{Journal of Geophysical Research: Planets}
\bvolume{124}(\bissue{3}),
\bfpage{752}--\blpage{778}
(\byear{2019})
\end{barticle}
\endbibitem

\end{thebibliography}
\restoregeometry
\end{document}